\documentclass[a4paper,twocolumn]{autart}

\usepackage{calrsfs,times,cite,xcolor}
\usepackage{graphicx,enumerate}
\usepackage{psfrag}
\usepackage{pstool}
\usepackage{amssymb}
\usepackage{amsmath}
\usepackage{calrsfs}
\usepackage{color}
\usepackage{epsfig,epstopdf,wrapfig}
\usepackage{paralist}
\usepackage{float,cuted}
\usepackage{multirow}
\usepackage{subcaption}

\newtheorem{theorem}{Theorem}
\newtheorem{remark}{Remark}

\newtheorem{lemma}{Lemma} 
\newtheorem{corollary}{Corollary} 
\newtheorem{assumption}{Assumption}
\newtheorem{proposition}{Proposition}
\newtheorem{problem}{Problem}

\newtheorem{problemprime}{Problem}

\def\tr{\mathop{\rm tr}\nolimits} 
\usepackage{amsmath}
\usepackage{accents}
\newlength{\dhatheight}

\usepackage{color}
\usepackage[normalem]{ulem}
\usepackage{soul} 
\soulregister\cite7
\soulregister\ref7
\soulregister\pageref7

\begin{document}
\begin{frontmatter}

\title{Coherent Equalization of Linear Quantum Systems}
\thanks[footnoteinfo]{This work was 
    supported by the Australian Research Council under the Discovery Projects funding scheme (project DP200102945).} 

\author[First]{V.~Ugrinovskii}\ead{v.ougrinovski@adfa.edu.au} \and
\author[Second]{M.~R.~James}\ead{Matthew.James@anu.edu.au}
\address[First]{School of Engineering and Technology, University of New
  South Wales 
  Canberra, Canberra, ACT, 2600, Australia}
\address[Second]{School of Engineering, The Australian National University,
  Canberra, ACT 2601, Australia} 

\begin{abstract}
This paper introduces a $H_\infty$-like methodology of
  coherent filtering for 
equalization of passive linear quantum systems to help mitigate 
degrading effects of quantum communication channels. For such systems, 
which include a wide range of linear quantum optical devices and signals,
we seek
to find a near optimal equalizing filter which is
itself a passive quantum system. The problem amounts to solving an 
optimization problem subject to constraints dictated by the requirement for
the equalizer to be physically realizable. By formulating these constraints 
in the frequency domain, we show that the problem admits a
convex $H_\infty$-like formulation. This allows us to derive a set of
  suboptimal coherent equalizers using $J$-spectral factorization. An
additional semidefinite relaxation      
combined with the Nevanlinna-Pick interpolation is shown to lead to a
tractable algorithm for the design of a near optimal
coherent equalizer. 
\end{abstract}

\volume{164, 111630}
\pubyear{2024}
\firstpage{}
\lastpage{}

\end{frontmatter}

\section{Introduction}

The main aim of this paper is to demonstrate an application of the
optimization paradigm to the derivation of 
coherent equalizers for a class of quantum systems that includes a wide range of linear quantum optical components and signals that may be used in quantum communication systems. 
Importantly, our design methodology is fully coherent, that is, it yields equalizers that are in the same class of quantum systems. 
Such equalizers 
are highly
desirable in quantum engineering since they do not require conventional
non-quantum measurement devices and hence are able to deliver
technological advantages of quantum information
processing. 
To be concrete, we 
focus on one type of the optimal coherent filtering
problem concerned with compensation of quantum signals
transmitted via a noisy quantum communication channel with a non-unity
frequency response, to help mitigate
distortions caused by the channel. In classical (i.e.,
  non-quantum) communications, such compensation of distortions is known as
  equalization.
Owing to the analogy with the
classical channel equalization, we call this
problem the \emph{quantum equalization problem}.

Optimization has proven to be an essential tool in the design of classical
communication and signal processing systems. The Wiener filtering
theory~\cite{Wiener-1949} is the best known demonstration of how degrading
effects of noise and the channel can be mitigated using optimization
techniques.  
While the Wiener's solution is elegant and tractable in the case of
stationary signals and perfectly known channels, additional properties and
requirements on the signal, the channel or the filter impose further
optimization constraints~\cite{DLS-2002,WBV-1999}. This is precisely the
situation encountered in the derivation of
a coherent  quantum filter, as such filter must
satisfy the fundamental constraints of \emph{physical realizability}, in
order to represent a valid quantum physical system 
in the class of systems considered;
see~\cite{JNP-2008,SP-2012,NY-2017,MP-2011} and references therein.  
\begin{figure}[t]
\begin{center}
\psfragfig[width=0.7\columnwidth]{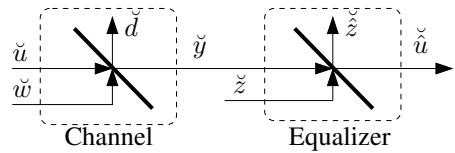}{ 
\psfrag{Quantum Channel}{\hspace{2ex}Channel}
\psfrag{Quantum Equalizer}{\hspace{2ex}Equalizer}
  \psfrag{u}{$\breve u$}
  \psfrag{w}{$\breve w$}
  \psfrag{hatw}{$\breve d$}
  \psfrag{z}{$\breve z$}
  \psfrag{hatu}{$\breve {\hat u}$}
  \psfrag{hatz}{$\breve {\hat z}$}
  \psfrag{y}{$\breve y$}}
  \caption{Example  quantum optical communication system consisting 
of a channel and equalizer, modelled by optical beamsplitters and  signals.
The message signal $\breve{u}$ is passed through the channel which may degrade the message,  resulting in the received signal $\breve{y}$. The equalizer near-optimally recovers the message in the mean square sense, producing an improved signal $\breve{\hat{u}}$.}
  \label{fig:bsplusbs}
\end{center}
\end{figure}
An example quantum optical communication system, shown in 
Fig.~\ref{fig:bsplusbs},
illustrates 
this situation. The signal (or message) being transmitted $\breve{u}$ passes through a lossy communication channel, modeled by the 
beamsplitter on the left  and a quantum noise signal $\breve{w}$.  The
received signal, $\breve{y}$, represents a degradation of the original
message $\breve{u}$.   
To compensate for this degradation, the received signal is
  passed through a filter designed to minimize the degradation error in the
  mean square sense, 
as in standard Wiener filtering. 
However, as we require the equalizing filter to be a quantum optical system,
this minimization is constrained by the above-mentioned physical realizability conditions, leading to an equalizer  design that involves an additional quantum noise $\breve{z}$, consistent with quantum mechanics. The equalizer (filter), shown in the right hand side of Fig.~\ref{fig:bsplusbs}, is represented by a beamsplitter and the additional quantum noise. In spite of the additional noise, the mean square error is reduced.  
The general
aim of the coherent equalization problem is to design quantum systems 
able to recover the transmitted message with high fidelity and  
which are realizable as physical quantum devices, hence the
  term coherent equalization. Using physical
  quantum devices as filters facilitates preserving information encoded in
  the quantum state of the system since detrimental effects of measurement
  back-action on the quantum state are avoided. In this paper we 
develop a procedure for the synthesis of transfer functions for such equalizing
filters.

Our focus is on the question whether
distortions introduced by a passive quantum communication channel can be
efficiently mitigated using another \emph{passive} quantum system acting as a
filter. Even restricted to passive filters, this question is
meaningful and sufficiently rich. Indeed, transfer functions
corresponding to passive coherent filters are easily implementable by
cascading quantum optical components such as beamsplitters, optical
cavities and phase shift devices~\cite{Nurdin-2010,NY-2017}. Therefore,
answering the 
question as to whether a (sub)optimal coherent equalizer can be obtained
within the class of passive systems enables synthesis 
of physical devices which solve coherent equalization problems. The paper
gives examples of such synthesis.   

The requirement for physical realizability
makes the task of finding an optimal coherent filter quite
nontrivial~\cite{VP-2013}. In~\cite{VP-2013}, this requirement led to nonconvex
constraints on the 
state-space matrices of the filter which prohibited obtaining a closed form
solution. In this paper, following~\cite{UJ2a,UJ2b}, we cast
the coherent  
equalization problem in the frequency domain. It turns out that in the frequency domain the physical
realizability constraints have a convenient structure.  
They can be partitioned so that the constraints on the
`key' variables which determine the filter performance can be separated
from the constraints on the ‘slack variables’ responsible for the physical
realizability of the filter. This leads us to adopt a two-step 
procedure for the design of coherent suboptimal filters which
  was originally proposed in~\cite{UJ2a}. In the first
step of this procedure, only some of the physical 
realizability constraints are retained, and the filter performance is
optimized over the `key' variables subject to these constraints. We
call this problem the auxiliary optimization problem. In the second step,
the remaining variables of the filter are computed to fulfill the
requirement of physical realizability. The rigorous justification of this
procedure is one of the original contributions of the paper. 

In contrast with the previous work~\cite{UJ2b,UJ2a,VP-2013}
concerned with developing coherent Wiener and Kalman filters,
we consider the problem in the vein of classical $H_\infty$
filtering~\cite{HSK-1999,Shaked-1990}. Also unlike the coherent $H_\infty$
control problem~\cite{JNP-2008,MP-2011a}, our
approach is concerned with 
minimization of the largest eigenvalue of the power spectrum density (PSD)
matrix of the equalization error. This allowed us to
formulate the aforementioned auxiliary optimization problem as 
a convex optimization problem whose constraints are frequency
dependent. 
This approach led to two contributions. Firstly, we
characterize the class of causal suboptimal coherent filters
in a manner similar to the Youla parameterization of $H_\infty$ suboptimal
controllers~\cite{ZDG-1996}, via the technique of 
$J$-spectral factorization~\cite{GGLD-1990,IO-1996}.
Secondly, we propose a Semidefinite Program (SDP) relaxation which
reduces the number of optimization constraints
to a finite number of constraints. Combined with the
method of Nevanlinna-Pick
interpolation~\cite{BGR-2013,DGK-1979,Kovalishina-1984} this gives
a tractable algorithm to obtain a physically realizable 
near optimal filter.

The optimization approach allows us to reveal some peculiar
features of coherent equalizers which set them apart from 
measurement-based filters. 
It turns out that,
unlike the classical equalization problem, the mean-square error between
the input and output fields of a linear quantum system may not always be
improved using a coherent linear equalizer. This is consistent
  with the earlier finding~\cite{UJ2a} that in the
simplest case when both the input field and the thermal noise field have one
degree of freedom and the channel is static, the coherent equalization is truly beneficial when 
the signal-to-noise ratio is below a certain threshold.
The
  paper relates the existence of such threshold 
to the question whether a certain
frequency dependent Linear Matrix Inequality (LMI) is feasible, as a sufficient
condition in the general case.

The paper is organized as follows. In the next section we present the
background on physically realizable open linear quantum systems. The
coherent equalization problem is also introduced in
Section~\ref{sec:equal-probl} where it is posed as an $H_\infty$-like
filtering problem subject to  
the physical realizability constraints. The 
justification of the two-step procedure for the design of coherent
suboptimal filters 
and the auxiliary optimization problem are presented in
Section~\ref{framework}. The relation between the
feasibility sets of this auxiliary problem and the corresponding classical
problem is also discussed in Section~\ref{framework}.  A complete
characterization of  
all suboptimal solutions for the auxiliary optimization problem is derived in
Section~\ref{feasible}. 
An alternative suboptimal solution to
this auxiliary problem via semidefinite programming and Nevanlinna-Pick
interpolation is presented in Section~\ref{semidef}. Section~\ref{examples}
presents two examples which illustrate these results. In the first example, the
results are applied to a single mode system
consisting of static components. The second example is an optical cavity system. For both examples, we
show how a suboptimal equalizer can be constructed via $J$-spectral
factorization, and also illustrate the semidefinite programming approach
undertaken in Section~\ref{semidef}. In the first example, we also
show that the bound on the performance delivered
via the $J$-spectral factorization approach is in fact tight
(i.e., it coincides with the solution obtained using
the Lagrange multiplier technique), and that an 
optimal equalizer can be obtained as a limit point of the set 
of suboptimal equalizers derived using the $J$-spectral factorization. This
example also
illustrates the threshold on the signal-to-noise ratio of the input fields
that arises due to the requirement of physical realizability. 
Conclusions and suggestions for future work are
  given in Section~\ref{Conclusions}.

\paragraph*{Notation}
For a collection of operators $\mathbf{a}_1$, \ldots, $\mathbf{a}_n$ in a
Hilbert space $\mathfrak{H}$, the notation $\mathrm{col}(\mathbf{a}_1,
\ldots, \mathbf{a}_n)$ denotes the column vector of operators obtained by
concatenating operators $\mathbf{a}_j$, i.e., the operator mapping
$\mathfrak{H}$ into the Cartesian product of $n$ copies of the space
$\mathfrak{H}$, $\mathfrak{H}^n$. For an operator
$\mathbf{a}:\mathfrak{H}\to \mathfrak{H}$, $\mathbf{a}^*$  denotes the
adjoint operator, and when   
$\mathbf{a}=\mathrm{col}(\mathbf{a}_1, \ldots, \mathbf{a}_n)$, 
$\mathbf{a}^\#$ denotes the column vector of adjoint operators,
$\mathbf{a}^\#=\mathrm{col}(\mathbf{a}_1^*, \ldots, \mathbf{a}_n^*)$,
$\mathbf{a}^T=(\mathbf{a}_1~\ldots~ \mathbf{a}_n)$ (i.e, the row of
operators), and $\mathbf{a}^\dagger = (\mathbf{a}^\#)^T$.  Also, we
will use the notation
$\breve{\mathbf{a}}=\mathrm{col}(\mathbf{a},\mathbf{a}^\#)=
\mathrm{col}(\mathbf{a}_1, \ldots, \mathbf{a}_n,\mathbf{a}_1^*, \ldots,
\mathbf{a}_n^*)$. $[\mathbf{a},\mathbf{b}]$ 
denotes the commutator of the operators $\mathbf{a},\mathbf{b}$ in
$\mathfrak{H}$,
$[\mathbf{a},\mathbf{b}]=\mathbf{a}\mathbf{b}-\mathbf{b}\mathbf{a}$. 
The quantum expectation of an operator $\mathbf{v}$ of a quantum system in
a state $\rho$ is denoted $\langle \mathbf{v}\rangle=\tr[\rho
 \mathbf{v}]$~\cite{Parthasarathy-2012}. 
For a complex number $a$, $a^*$ is its complex conjugate and
for a matrix $A=(A_{ij})$, $A^\#$, $A^T$, $A^\dagger$ denote, respectively,
the matrix of complex conjugates $(A_{ij}^*)$, the transpose matrix and the
Hermitian adjoint matrix. 
$I$ is the identity matrix, and 
$ 
J=
  \left[
  \begin{array}{rr}
    I & 0 \\ 0 & -I
  \end{array}
  \right].
$
We will also write $I_n$ when we need to specify that this
is the $n\times n$ identity matrix.
For two complex matrices $X_-$, $X_+$, we write 
$
  \Delta(X_-,X_+)=
  \left[
    \begin{array}{cc} X_- & X_+ \\ X_+^\# & X_-^\# 
    \end{array}
  \right].
$
When $X_-=X_-(s)$, $X_+=X_+(s)$ are complex transfer function matrices, the
stacking operation defines the transfer function matrix
$
  \Delta(X_-(s),X_+(s))=
  \left[
    \begin{array}{cc} X_-(s) & X_+(s) \\ (X_+(s^*))^\# & (X_-(s^*))^\# 
    \end{array}
  \right].
$
For a transfer
function matrix $X(s)$, $X(s)^H$ denotes its Hermitian para-conjugate,
$X(s)^H=X(-s^*)^\dagger$. Clearly, for a complex matrix $X$, $X^H=X^\dagger$. 
When the matrix $X$ is Hermitian, $\boldsymbol{\sigma}(X)$ is the largest eigenvalue
of $X$. For a transfer function $X(s)$ in the Hardy space $H_\infty$ of
matrix-valued functions which are analytic in the open right
half-plane $\mathrm{Re}s>0$ and are bounded on the imaginary axis,
$\|X\|_\infty$ denotes its $H_\infty$ norm, $\|X\|_\infty=\sup_{\mathrm{Re}s>0}\|X(s)\|=\mathrm{ess}\sup_{\omega\in\mathbf{R}}\|X(i\omega)\|$,
where $\|\cdot\|$ is the induced 2-norm of a matrix~\cite{ZDG-1996}.

\section{Coherent equalization problem for linear quantum communication
  systems}\label{sec:equal-probl}

\subsection{Linear Open Quantum Systems}
\label{sec:new-open-systems}

In this paper we use a class of open quantum systems to describe quantum
signals, channels and devices. Open quantum systems are widely used, e.g.,
\cite{GZ-2000,WM-2008,WM-2009}.
The particular class of open quantum systems we use involves systems for
which creation and annihilation operators evolve linearly in the Heisenberg
picture. This class of linear open quantum systems includes a wide range of
linear quantum optical systems and signals~\cite{NY-2017}.

Perhaps the simplest example of a linear open quantum system is a cavity
with mode $\mathbf{a}$ coupled to a vacuum (input) field $\mathbf{b}(t)$
with coupling 
constant $2\kappa$. 
Here, the 
mode annihilation operator $\mathbf{a}$ evolves unitarily, $\mathbf{a}(t) =
\mathbf{U}(t)^\ast \mathbf{a} \mathbf{U} (t)$, and the output field is
given by $\mathbf{b}_{out}(t) = \mathbf{U}(t)^\ast \mathbf{b}(t) \mathbf{U}
(t)$; $\mathbf{U}(t)$ is the unitary operator as described in~\cite{GZ-2000}. Together, the mode operator
and the output field satisfy the  linear Heisenberg
equations~\cite{GJN-2010,SP-2012}  
\begin{eqnarray}
&&\dot {\mathbf{a}}(t) = -\kappa \mathbf{a}(t) - \sqrt{2\kappa}
  \mathbf{b}(t), \nonumber \\ 
&&\mathbf{b}_{out}(t) = \sqrt{2\kappa} \mathbf{a}(t) + \mathbf{b}(t).
\label{eq:new-cavity-simple}
 \end{eqnarray}
Note that while the unitary $\mathbf{U}(t)$ and annihilation
$\mathbf{a}(t)$ operators evolve linearly, other operators, such as the
number operator $\mathbf{n}=\mathbf{a}^\ast \mathbf{a}$ may have nonlinear
evolution. 

More generally, the class of linear open quantum systems we employ in this paper is specified in terms of $m$ harmonic modes $\mathbf{a}_j,  j=1,\ldots m$,
driven by $n$ quantum input  
fields (signals) $\mathbf{b}_k(t), k=1,\ldots n$, ~\cite{HP-1984,GJN-2010,ZJ-2013}. 
We use the vector notation 
$\mathbf{a}=\mathrm{col}(\mathbf{a}_1, \ldots \mathbf{a}_m)$,
and $\mathbf{a}^\#=\mathrm{col}(\mathbf{a}_1^*,\ldots, \mathbf{a}_m^*)$ for their adjoint operators; combining these 
 we write
 $\breve{\mathbf{a}}    =  \mathrm{col}( \mathbf{a}, \mathbf{a}^\#)$.
The commutation relations are $[\mathbf{\breve{a}},\mathbf{\breve{a}}^\dagger]=J$.

 The
quantum input fields are represented as annihilation and creation operators
$\mathbf{b}(t)=\mathrm{col}(\mathbf{b}_1(t), \ldots, \mathbf{b}_n(t))$, 
$\mathbf{b}^\#(t)=\mathrm{col}(\mathbf{b}^\#_1(t), \ldots,
\mathbf{b}^\#_n(t))$,
satisfying canonical commutation relations
$  [\breve{\mathbf{b}}(t),\breve{\mathbf{b}}^\dagger(t')]=J\delta(t-t')$; 
here $\delta(t)$ is the delta function.
 When the input
fields are in a Gaussian state with zero mean, which is the situation
considered in this paper, these random processes can be regarded as
stationary quantum Gaussian white noise processes with zero mean (i.e., $\langle
\breve{\mathbf{b}}(t)\rangle =0$) and the correlation function
\begin{eqnarray}
  \label{eq:8}
  \langle\breve{\mathbf{b}}(t)\breve{\mathbf{b}}^\dagger(t')\rangle=F_{\mathbf{b}}
\delta(t-t'),
  \quad F_{\mathbf{b}}\triangleq
 \left[
    \begin{array}{cc}
I+\Sigma_{\mathbf{b}}^T & \Pi_{\mathbf{b}} \\
 \Pi_{\mathbf{b}}^\dagger & \Sigma_{\mathbf{b}}
\end{array}\right].
\end{eqnarray}
The matrix $F_{\mathbf{b}}$ symbolizes intensity of
the process $\breve{\mathbf{b}}$; $\Sigma_{\mathbf{b}}$ is a nonnegative definite
complex Hermitian matrix, 
$\Sigma_{\mathbf{b}}^\dagger=\Sigma_{\mathbf{b}}$, and $\Pi_{\mathbf{b}}$ 
is a complex symmetric matrix, $\Pi_{\mathbf{b}}^T=\Pi_{\mathbf{b}}$. In
this paper, we will consider linear quantum systems in thermal state, therefore
it will always be assumed that $\Pi_{\mathbf{b}}=0$.

In the Heisenberg picture, the system dynamics are given by
\begin{eqnarray}
  \label{dyn}
  \dot{\breve{\mathbf{a}}}(t)&=&\breve A \breve{\mathbf{a}}(t)+  \breve B
                              \breve{\mathbf{b}}(t), \nonumber \\
  \breve{\mathbf{y}}(t)&=&\breve C \breve{\mathbf{a}}(t)+ \breve D
                           \breve{\mathbf{b}}(t).
\end{eqnarray}
Here, 
$\breve{\mathbf{y}}=\mathrm{col}(\mathbf{y},\mathbf{y}^\#)$ denotes the
output field of the system that carries away information about the system
interacting with the input field $\breve{\mathbf{b}}$; the vectors of operators 
$\mathbf{y}=\mathcal{col}(\mathbf{y}_1, \ldots, \mathbf{y}_{n_b})$, 
$\mathbf{y}^\#=\mathcal{col}(\mathbf{y}_1^*, \ldots, \mathbf{y}_{n_b}^*)$, 
have the same dimension $n_b$ as the dimension of the vectors
$\mathbf{b}$, $\mathbf{b}^\#$ of the input field. The matrices
$\breve{A}$, $\breve{B}$, $\breve{C}$, $\breve{D}$ are complex matrices\footnote{The problem
of finding parameters of the system~(\ref{dyn}) is treated in the large body
  of literature on quantum system identification.}
partitioned in accordance with the structure of the vectors
of operators $\breve{\mathbf{a}}$, $\breve{\mathbf{b}}$, as 
\begin{eqnarray*}
  \label{eq:4}
\breve{A}&=&\Delta(A_-,A_+), \quad   
\breve{B}=\Delta(B_-,B_+), \\
\breve{C}&=&\Delta(C_-,C_+), \quad   
             \breve{D}=\Delta(D_-,D_+).
\end{eqnarray*}
A detailed discussion of open linear quantum systems can be found 
in~\cite{JG-2010,GJN-2010,JNP-2008,ZJ-2013}. The cavity equations
(\ref{eq:new-cavity-simple}) are a special case of (\ref{dyn}).  

In the subsequent
sections we will consider passive linear quantum systems. For such systems,  
$A_+=0$, $B_+=0$, $C_+=0$, $D_+=0$, i.e., the evolution of the `annihilation
part' of the system variable $\breve{\mathbf{a}}$ is governed 
  only by the annihilation operators $\mathbf{b}(t)$ of the input field,
  and the `creation' part of $\breve{\mathbf{a}}$ 
  is driven by the creation operators $\mathbf{b}^\#(t)$~\cite{JG-2010}. In
  this case, the matrices $\breve A$, $\breve B$, $\breve C$, $\breve D$
  are block diagonal.

For a quantum stochastic differential equation of the form
(\ref{dyn}) to describe evolution of quantum physical system in the
Heisenberg picture, its coefficients must satisfy certain
additional conditions~\cite{SP-2012,JNP-2008,MP-2011}. These conditions,
known as 
the \emph{physical realizability conditions}, ensure that the oscillator
variables $\mathbf{a}_j(t)$ and the output field operators
$\mathbf{y}_j(t)$ defined by equation (\ref{dyn}) evolve unitarily. We next describe the physical realizability  conditions in the frequency domain.

In the frequency domain, the input-output map
defined by system~(\ref{dyn})
is expressed in terms of the $n_b\times
n_b$ transfer function
\[ 
  \Gamma(s)= \breve{C}(sI_{2m}-\breve{A})^{-1}\breve{B}+\breve{D},
\] 
relating the bilateral Laplace transforms of 
$\breve{\mathbf{y}}(t)$ and $\breve{\mathbf{b}}(t)$~\cite{GJN-2010,ZJ-2013}. 
According to the next lemma, physical realizability of
the linear quantum system (\ref{dyn}) dictates that the transfer function
$\Gamma(s)$ must be $J$-symplectic\footnote{A
  transfer function matrix $\Gamma(s)$ is
  $J$-symplectic if $\Gamma J\Gamma^H=J$~\cite{GJN-2010}. Such transfer
  functions are also known as 
  $J$-unitary~\cite{BGR-2013} and $(J,J)$-unitary~\cite{SP-2012}.};
see~\cite{SP-2012,GJN-2010,BGR-2013}. The lemma is a straightforward
combination of Theorem~4 in~\cite{SP-2012} and Theorem~6.1.1
in~\cite{BGR-2013}. It requires the following assumption about the linear
quantum system~(\ref{dyn}).

\begin{assumption}
  \label{A1}
The pair $(\breve A,\breve B)$ is controllable, and the
pair $(\breve A,\breve C)$ is observable.
\end{assumption}

\begin{lemma}\label{L.pr=unitary}
Suppose Assumption~\ref{A1} is satisfied. Then the 
following conditions are equivalent:
\begin{enumerate}[(a)]
\item 
The linear quantum system~(\ref{dyn}) is physically realizable;
\item 
$\breve D=\Delta(S,0)$ where $S$ is a constant unitary matrix, and
\begin{equation}
  \label{eq:1}
\Gamma(s)J\Gamma(s)^H=\Gamma(s)^H J \Gamma(s)=J; 
\end{equation}
\item 
$\breve D=\Delta(S,0)$ where $S$ is a constant unitary matrix, and
\begin{equation}
  \label{eq:1w}
\Gamma(i\omega)J\Gamma(i\omega)^\dagger=\Gamma(i\omega)^\dagger J
\Gamma(i\omega)=J.  
\end{equation}
\end{enumerate}
\end{lemma}

When the system (\ref{dyn}) is a passive (annihilation only) system, its
transfer function $\Gamma(s)$ is block-diagonal~\cite{GJN-2010}
\begin{eqnarray}
  \label{eq:31}
&&  \Gamma(s)=  \left[
    \begin{array}{cc}
      G(s)& 0 \\
      0 & G(s^*)^\#
    \end{array}
  \right], 
\end{eqnarray}
where $G(s)$ is the transfer function of the annihilation part of the
system, $G(s)=C_-(sI-A_-)^{-1}B_-+S$. Assumption~\ref{A1} reduces to the
assumption 
that $(A_-,B_-)$ and $(A_-,C_-)$ are controllable and observable,
respectively. It then follows from this assumption that the matrices $A_-$
and $\breve A=\Delta(A_-,0)$ are Hurwitz and that $G(s)$ and $\Gamma(s)$
in~(\ref{eq:31}) are stable rational proper transfer functions. 
Furthermore, the frequency domain physical realizability
relations~(\ref{eq:1}),~(\ref{eq:1w})  reduce to the condition that $D_-=S$
is a unitary matrix, $G(s)$
is paraunitary and $G(i\omega)$ is unitary
(also see \cite{MP-2011}), 
\begin{eqnarray}
  \label{eq:32}
  G(s)^H G(s)=G(s)G(s)^H=I, \\
  G(i\omega)^\dagger G(i\omega)=G(i\omega)G(i\omega)^\dagger=I.
  \label{eq:32.w}
\end{eqnarray}
Then $G(i\omega)$ is bounded at infinity and
analytic on the entire closed imaginary axis~\cite[Lemma~2]{Youla-1961}.

\subsection{A quantum communication system}
 
\begin{figure}[t]
\begin{center}
\psfragfig[width=0.7\columnwidth]{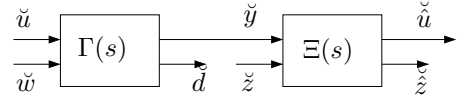}{ 
\psfrag{Quantum}{}
\psfrag{channel}{$\Gamma (s)$}
\psfrag{equalizer}{$\Xi(s)$}
  \psfrag{+}{$+$}
  \psfrag{-}{$-$}
  \psfrag{e}{$\breve e$}
  \psfrag{b}{$\breve u$}
  \psfrag{w}{$\breve w$}
  \psfrag{what}{$\breve d$}
  \psfrag{z}{$\breve z$}
  \psfrag{bhat}{$\breve {\hat u}$}
  \psfrag{zhat}{$\breve {\hat z}$}
  \psfrag{y}{$\breve y$}}
  \caption{A general quantum communication system. The transfer function
    $\Gamma(s)$ represents the channel, and $\Xi(s)$ represents an
    equalizing filter.} 
  \label{fig:general}
\end{center}
\end{figure}

We consider a general setup consisting of a linear
quantum system representing a communication channel and a
second linear quantum system acting as an equalizer, as shown in
Fig.~\ref{fig:general}.  
The $n$-dimensional input vector field $\breve u$ plays the role of a
signal carrying a message  
transmitted through the channel, and the $n_w$-dimensional vector of
operators $\breve w$ is comprised of 
operators describing the environment as well as noises introduced by the
routing hardware such as beamsplitters, etc. In what follows it is assumed
that these operators commute,
$  [\breve u,\breve w]=0,
$ 
and the system is in a Gaussian thermal state, and $\langle \breve u(t)\rangle=0$, $\langle\breve w(t)\rangle=0$. Furthermore, it is assumed
that the input fields $\breve u$ and $\breve w$ are 
not correlated, $\langle \breve u (t) \breve w^\dagger (t')\rangle=0$.

To represent the communication channel
as a linear quantum system, the annihilation and
creation parts of $\breve u$, $\breve w$ are stacked together to form
the vectors of operators    
$\mathbf{b}=\mathrm{col}(u,w)$ and $\mathbf{b}^\#=\mathrm{col}(u^\#,w^\#)$,
which are then combined into the vector $\breve{\mathbf{b}}$. 
This combined 
vector of input operators $\breve{\mathbf{b}}$ is applied to a linear
quantum system with the transfer function $\Gamma(s)$ which
  represents the communication channel. The annihilation and
creation operators of the output field of this system form the vector
$\breve{\mathbf{y}}=\mathrm{col}(y,d,y^\#,d^\#)$. The dimensions of the
annihilation operators $y$ and
$d$ (respectively, creation operators $y^\#$ and $d^\#$) of the output
field are $n$ and $n_w$, respectively.

Introduce the partition of the
transfer function $G(s)$ compatible with the partition  
$\mathbf{b}=\mathrm{col}(u,w)$, $\mathbf{y}=\mathrm{col}(y,d)$,
\begin{eqnarray}
   \label{eq:98}
  G(s)&=&
  \left[
    \begin{array}{cc}
G_{11}(s) & G_{12}(s)\\
G_{21}(s) & G_{22}(s)\\
    \end{array}
  \right].
\end{eqnarray}
Using this partition, the physical realizability condition~(\ref{eq:32})
reduces to the identities  
\begin{subequations}
\label{eq:35}
\begin{align}
  \label{eq:9G}
&
G_{11}(s)G_{11}(s)^H+G_{12}(s)G_{12}(s)^H =I, \\
&
G_{11}(s)G_{21}(s)^H+G_{12}(s)G_{22}(s)^H =0, \label{eq:10G}
\\
& 
G_{21}(s)G_{21}(s)^H+G_{22}(s)G_{22}(s)^H =I. \label{eq:11G}
\end{align}  
\end{subequations}

Recall a frequency domain relationship
between power spectrum densities of the input and the stationary output
fields of the linear quantum system~(\ref{dyn}). Since we focus on passive
systems, we restrict attention to the autocorrelation matrix of 
$\mathbf{y}$, 
$R_{\mathbf{y}}(t)=\langle
\mathbf{y}(t)\mathbf{y}^\dagger(0)\rangle$. 
The corresponding power spectrum density (PSD) matrix $P_{\mathbf{y}}(s)$
is the bilateral Laplace transform of $R_{\mathbf{y}}(t)$~\cite{ZJ-2013}. 
It was shown in~\cite{ZJ-2013} that since the matrix $\breve A$ is
Hurwitz, 
it holds that
\begin{equation}
  P_{\mathbf{y}}(i\omega)=
G(i\omega) (I+\Sigma_{\mathbf{b}}^T)G(i\omega)^\dagger.  
  \label{PSD}  
\end{equation} 

A coherent equalizer is another linear quantum system $\Xi(s)$ 
which takes the components $\breve y=\mathrm{col}(y,y^\#)$ of the output
$\breve{\mathbf{y}}$ as one of its inputs. Its second 
input $\breve{z}(t)=\mathrm{col}(z(t),z^\#(t))$ in Fig.~\ref{fig:general}
is comprised of $n_z$ annihilation and creation operators of the auxiliary noise
input field introduced into the filter model to make it physically
realizable~\cite{JNP-2008,VP-2011}. 
For simplicity, we assume that the filter environment is in a Gaussian vacuum
state; that is, the filter's quantum noise process $\breve{z}(t)$ has zero
mean, $\langle \breve z(t)\rangle =0$, and the correlation function
$  \langle
     \breve z(t) 
     \breve{z}^\dagger(t')\rangle=
  \left[
    \begin{array}{cc}
I_{n_z} & 0 \\
0 & 0
\end{array}\right]\delta(t-t'),
$ 
i.e., $\Sigma_z=0$, $\Pi_z=0$. The operator $\breve{z}$ commutes with
$\breve{u}$ and $\breve{w}$.

The input into the equalizer,
$\breve{\mathbf{b}}_{\textrm{eq}}=\mathrm{col}(\mathbf{b}_{\textrm{eq}},\mathbf{b}_{\textrm{eq}}^\#)$, 
combines the output $\breve y$ of the channel and the filter environment
noise $\breve z$, so that $\mathbf{b}_{\textrm{eq}}=\mathrm{col}(y,z)$. Its
output is $\breve {\mathbf{y}}_{\textrm{eq}}
=\mathrm{col}(\mathbf{y}_{\textrm{eq}},\mathbf{y}_{\textrm{eq}}^\#)$, and
each of the vectors of operators
$\mathbf{y}_{\textrm{eq}},\mathbf{y}_{\textrm{eq}}^\#$ can be partitioned   
into operator vectors whose dimensions match the dimensions of
$y$ and $z$, respectively: 
$\mathbf{y}_{\textrm{eq}}=\mathrm{col}(\hat u,\hat z)$, 
$\mathbf{y}_{\textrm{eq}}^\#=\mathrm{col}(\hat u^\#,\hat z^\#)$. We
designate the 
first component of these partitions, namely $\hat u$ (respectively $\hat
u^\#$), as the output field of the equalizing filter.

   Since the focus of this paper is on passive coherent
filters\footnote{Some results on active equalization 
  can be found
  in~\cite{UJ2b}.}, from now only
the filters of the form $\Xi(s)=\Delta(H(s),0)$ will be considered, where
$H(s)$ is an $(n+n_z)\times (n+n_z)$ transfer function. 
According to Lemma~\ref{L.pr=unitary}, physical 
realizability of the filter  
requires that $H(s)$ must be a paraunitary transfer function matrix and the
matrix $H(i\omega)$ must be unitary; cf.~(\ref{eq:32}),~(\ref{eq:32.w}): 
\begin{equation}
  \label{eq:60}
  H(s)H(s)^H=I, \quad H(i\omega)H(i\omega)^\dagger=I. 
\end{equation}
The set of equalizers $\Xi(s)=\Delta(H(s),0)$, where 
$H(s)$ satisfies~(\ref{eq:60})
will be denoted $\mathcal{H}_p$. 

The transfer function matrix $H(s)$ can be further partitioned into
the blocks compatible with dimensions of the filter inputs
$\mathrm{col}(y,z)$ and outputs $\mathrm{col}(\hat u, \hat z)$:   
\begin{eqnarray}
  \label{eq:98a}
  H(s)=
  \left[
    \begin{array}{cc}
H_{11}(s) & H_{12}(s)\\H_{21}(s) & H_{22}(s)
    \end{array}
  \right]. \qquad 
\end{eqnarray} 
Using this partition, the condition (\ref{eq:60}) can be expanded into
conditions of the form~(\ref{eq:35}) which provide an explicit set of
constraints imposed on the transfer functions of each of the filter
channels by the requirement for physical realizability
\begin{subequations}
\label{eq:37}
\begin{align}
  \label{eq:9p}
&
H_{11}(s)H_{11}(s)^H+H_{12}(s)H_{12}(s)^H =I, \\
&
H_{11}(s)H_{21}(s)^H+H_{12}(s)H_{22}(s)^H =0, \label{eq:10p}
\\
& 
H_{21}(s)H_{21}(s)^H+H_{22}(s)H_{22}(s)^H =I. \label{eq:11p}
\end{align}  
\end{subequations}

\subsection{The coherent equalization problem}\label{sec:eqprob-loose}

Let $e(t)$ be the difference between the channel
input and the filter output fields, $e(t)=\hat u(t)-u(t)$. We refer to
$e(t)$ as the equalization error of the filter $\Xi(s)$. Let $P_e(i\omega)$
denote the Fourier transform of the autocorrelation 
matrix $R_e(t)= \langle e(t)e(0)^\dagger \rangle$. $P_e(i\omega)$
represents the power spectrum density of the difference between the channel
input and the filter output fields. \emph{The coherent equalization problem} in
this paper is to obtain a physically realizable passive filter transfer function
$\Xi$ which minimizes (exactly or approximately) the largest eigenvalue
of $P_e(i\omega)$:
\begin{eqnarray}
  \label{eq:6}
\Xi=\mathrm{arg} \inf_{\Xi}\sup_\omega
  \boldsymbol{\sigma}(P_e(i\omega)).
\end{eqnarray}

\subsection{The formal problem statement}\label{statement}

Let $e(t)=\hat u(t)-u(t)$ be the equalization error introduced in
Section~\ref{sec:eqprob-loose}. We then write that $\breve
e(t)=\mathrm{col}(e(t),e^\#(t))$. The transfer
function from the combined `input plus channel and filter environment'
field $\breve{\mathbf{v}}=\mathrm{col}(u,w,z,u^\#,w^\#,z^\#)$ to
$\breve e$ is obtained by interconnecting the passive channel and passive
filter systems as shown in Fig.~\ref{fig:general},    
\begin{eqnarray}
  E(s)&=&\Delta(E_-(s), 0), \nonumber \\
E_-(s)&\triangleq &\left[ \begin{array}{c|c|c}
    H_{11}(s) G_{11}(s) -I ~&~ H_{11}(s) G_{12}(s)~  &~ H_{12}(s)
              \end{array}\right].\qquad
\label{eq:34} 
\end{eqnarray}
Using this transfer function and~(\ref{PSD}), the Fourier transform  of the
autocorrelation 
matrix of the equalization error
$  R_e(t)= \langle e(t)e(0)^\dagger \rangle$
can be expressed as  
\begin{eqnarray}
  P_{e}(i\omega)&=&
    \left[
    \begin{array}{cc}
      E_-(i\omega)& 0
    \end{array}
    \right] F_{\mathbf{v}}     \left[
    \begin{array}{c}
      E_-(i\omega)^\dagger \\ 0
    \end{array}
    \right].
\label{eq:39}
  \end{eqnarray}
Here $F_{\mathbf{v}}$ is the intensity matrix of the noise process
$\breve{\mathbf{v}}$ when the system is in a thermal quantum state  
(cf.~(\ref{eq:8})),
\begin{eqnarray}  
F_{\mathbf{v}}&=&\left[ \begin{array}{ccc|ccc}
I+\Sigma_u^T & 0 & 0 & 0 & 0 & 0 \\
0 & I+\Sigma_w^T & 0 & 0 & 0 & 0 \\
0 & 0 & ~I~ & 0 & 0 & 0 \\ \hline
0 & 0 & 0 & \Sigma_u & 0 & 0 \\
0 & 0 & 0 & 0 & ~\Sigma_w~ & 0 \\
0 & 0 & 0 & 0 & 0 & 0
\end{array}
\right].
\label{eq:38.F}
\end{eqnarray}  
Since $F_{\mathbf{v}}^\dagger=F_{\mathbf{v}}$, $P_{e}(i\omega)$ is an $n\times n$ Hermitian matrix
where $n$ is the dimension of the `channel input' $u$. Hence the eigenvalues 
of $P_{e}(i\omega)$ are real.  

Using (\ref{eq:34}) and~(\ref{eq:38.F}) an
explicit expression for $P_e(i\omega)$ can be obtained~\cite{UJ2a},
\begin{eqnarray}
\label{eq:121}
\lefteqn{P_e(i\omega)} && \nonumber \\
&=&
(H_{11}(i\omega)G_{11}(i\omega)-I)(I+\Sigma_u^T)(G_{11}(i\omega)^\dagger 
H_{11}(i\omega)^\dagger -I) 
\nonumber \\
&+&
H_{11}(i\omega)G_{12}(i\omega)(I+\Sigma_w^T)G_{12}(i\omega)^\dagger H_{11}(i\omega)^\dagger \nonumber \\
&+&
    H_{12}(i\omega)H_{12}(i\omega)^\dagger.
\end{eqnarray}
Taking advantage of the properties~(\ref{eq:9G}) and~(\ref{eq:9p}) due to
passivity of the channel and filter transfer functions, this expression
can be simplified:  
\begin{eqnarray}
\label{eq:121p}
\lefteqn{P_{e}(i\omega)=H_{11}(i\omega) \Psi(i\omega) H_{11}(i\omega)^\dagger} && \nonumber  \\
&-&
    H_{11}(i\omega)G_{11}(i\omega)(I+\Sigma_u^T)-(I+\Sigma_u^T)G_{11}(i\omega)^\dagger H_{11}(i\omega)^\dagger
\nonumber \\
&+&   \Sigma_u^T+2I,
\end{eqnarray}
where we let
\begin{equation}
  \label{eq:47}
\Psi(s)\triangleq G_{11}(s)\Sigma_u^TG_{11}(s)^H
       +G_{12}(s)\Sigma_w^TG_{12}(s)^H.   
\end{equation}
In the sequel, we will also make use of the $n\times n$ matrix 
\begin{eqnarray}
  \label{eq:59}
 \lefteqn{ P_e(s)=
  \left[
    \begin{array}{cc}
      E_-(s)& 0
    \end{array}
  \right] F_{\mathbf{v}}   \left[
    \begin{array}{c}
      E_-(s)^H \\ 0
    \end{array}
  \right]} && \nonumber \\
&=&H_{11}(s) \Psi(s) H_{11}(s)^H -H_{11}(s)G_{11}(s)(I+\Sigma_u^T)
\nonumber \\
&-&
    (I+\Sigma_u^T)G_{11}(s)^H H_{11}(s)^H
+   \Sigma_u^T+2I.
\end{eqnarray}
Again, this expression is obtained using the identities~(\ref{eq:9G})
and~(\ref{eq:9p}). We will also write
$P_e(s,H)$ when we need to stress that the
expression for $P_e(s)$ corresponds to a specific filter
$\Xi(s)=\Delta(H(s),0)$.
 
We now present a formal statement of the problem of
coherent passive equalization posed in
Section~\ref{sec:eqprob-loose}.
 
\begin{problem}\label{P1}
The guaranteed cost passive equalization problem is to obtain a transfer
function matrix $\Xi(s)=\Delta(H(s),0)\in 
\mathcal{H}_p$ which ensures a desired
bound on the power spectrum density of the equalization error. That is, given
$\gamma>0$, obtain $\Xi(s)=\Delta(H(s),0)\in \mathcal{H}_p$ such that
\begin{equation}
  P_e(i\omega)<\gamma^2 I_n \quad \forall\omega\in\bar{\mathbf{R}},    \label{eq:6'.sub}
\end{equation}
here $\bar{\mathbf{R}}$ denotes the closed real axis: $\bar{\mathbf{R}}=\mathbf{R}\cup\{\pm\infty\}$.   
The optimal passive equalization problem is to minimize the
bound~(\ref{eq:6'.sub}) in the
class of filters $\mathcal{H}_p$:
\begin{eqnarray}
  \label{eq:6'}
&&\gamma_\circ\triangleq \inf \gamma  \mbox{ subject to
   $\gamma>0$ and (\ref{eq:6'.sub})}.
\end{eqnarray}
\end{problem}

In~Problem~\ref{P1} we tacitly replaced optimization of
$\sup_\omega\boldsymbol{\sigma}(P_e(i\omega))$ with~(\ref{eq:6'}). The two
problems are equivalent. Indeed, given $\gamma>0$, define the set
\[
\mathcal{H}_\gamma=\{H(s)\colon \sup_\omega\boldsymbol{\sigma}(P_e(i\omega,H))<\gamma^2, H(s)H(s)^H=I\}.
\]
\begin{lemma}\label{L.eq=Hinf}
  \begin{equation}
    \label{eq:41}
    \gamma_\circ=\bar\gamma\triangleq \inf\{\gamma>0\colon
    \mathcal{H}_\gamma\neq \emptyset\}. 
  \end{equation}
\end{lemma}

\emph{Proof: }
From the definition of $\bar\gamma$, there exists a sequence
$\{\gamma_k\}\subset 
\{\gamma>0\colon \mathcal{H}_\gamma\neq 
\emptyset\}$ such that $\gamma_k\ge \bar\gamma$ and
$\lim_{k\to\infty}\gamma_k= \bar\gamma$. That is, for 
any $\epsilon>0$, one can choose a sufficiently large $k$ so that
$\gamma_k<\bar\gamma+\epsilon$. Also, since $\mathcal{H}_{\gamma_k}\neq
\emptyset$, there exists a passive $H_k(s)$ such that
$\sup_\omega\boldsymbol{\sigma}(P_e(i\omega,H_k))<\gamma_k^2$. Consequently, 
$P_e(i\omega,H_k)<\gamma_k^2I_n$ for any $\omega\in\bar{\mathbf{R}}$, therefore
$ \gamma_\circ \le \gamma_k<\bar\gamma+\epsilon$.
Letting $\epsilon\to 0$ implies that
$\gamma_\circ\le\bar\gamma$. 

Conversely, according to the definition of $\gamma_\circ$, there exists a sequence of constants $\{\gamma_l'\}$,
$\gamma_l'\ge \gamma_\circ $, which 
converges to $\gamma_\circ$ and such that for each $\gamma_l'$ there exists
a physically realizable $H_l(s)$ such that
$P_e(i\omega,H_l)<(\gamma_l')^2I_n$ for any $\omega$. Then
$\sup_\omega\boldsymbol{\sigma}(P_e(i\omega,H_l))\le
(\gamma_l')^2<(\gamma_l'+\epsilon)^2$, where $\epsilon>0$ is an arbitrarily
small constant. Thus, $\mathcal{H}_{\gamma_l'+\epsilon}\neq \emptyset$,
which means that $\bar\gamma\le \gamma_l'+\epsilon$.
Letting $l'\to\infty$, $\epsilon\to 0$ leads to the conclusion that
$\bar\gamma\le \gamma_\circ$. Thus, $\bar\gamma= \gamma_\circ$.
\hfill$\Box$

Lemma~\ref{L.eq=Hinf} indicates that Problem~\ref{P1} is analogous to the
classical $H_\infty$ filtering problem~\cite{HSK-1999}. 
However, instead of the singular value of the disturbance-to-error transfer
function, we seek to optimize the largest eigenvalue of the PSD function
$P_e$. Importantly, Problem~\ref{P1}
belongs  to the class of \emph{constrained} optimization problems since the
class of 
admissible filters is restricted to physically realizable passive
filters. 

\section{The framework for solving Problem~\ref{P1}}\label{framework}

\subsection{The procedure for the synthesis of coherent
  equalizers}
\label{sec:two-step} 

The expression for $P_{e}$ obtained in~(\ref{eq:121p})
depends only on 
$H_{11}$ and does not depend explicitly on other blocks 
of the matrix $H$. 
Therefore we adopt a two-step procedure
to solve Problem~\ref{P1} which was originally proposed in~\cite{UJ2a}. In the first step of this
procedure, the power spectrum density of the equalization error will be
optimized with respect to $H_{11}(s)$ subject to 
some of the constraints implied by the paraunitarity of $H$. 
Next, the blocks $H_{12}(s)$, $H_{21}(s)$, $H_{22}(s)$ of the equalizer
transfer function will be computed to fulfill the constraint~(\ref{eq:37}). However, \cite{UJ2a} did not explain how causal $H_{12}(s)$,
  $H_{21}(s)$, $H_{22}(s)$ can be computed. This problem is solved in this
  section. For this, we   
recall the notion of spectral
factors of a rational para-Hermitian\footnote{A rational transfer
  function matrix $X(s)$ is para-Hermitian if $X(s)^H=X(s)$.}
transfer function matrix~\cite{Youla-1961}. 

\begin{lemma}[Youla, Theorem~2 of~\cite{Youla-1961}]\label{Youla.T2}
Suppose a rational
para-Hermitian $n\times n$ transfer function matrix $X(s)$ is positive
semidefinite on the imaginary axis, $X(i\omega)\ge 0$, and has normal rank\footnote{A non-negative
  integer $r$ is the normal rank of a rational function $X(s)$ if (a)
  $X$ has at least one subminor of order $r$ which does not vanish
  identically,  and (b) all minors of order greater than $r$ vanish
  identically~\cite{Youla-1961}.} $r$, $r\le n$. Then the following statements hold. 
\begin{enumerate}[(a)]
\item There exists an $r\times n$ rational matrix $N(s)$
such that $X(s)=N(s)^HN(s)$. $N(s)$ is a spectral factor of $X(s)$.
\item $N(s)$ and its right inverse $N^{-1}(s)$ are both analytic in the
  open right half-plane $\mathrm{Re}s>0$.
\item
  $N(s)$ is unique up to a constant unitary $r\times r$ matrix multiplier
  on the left; i.e., if $N_1(s)$ also satisfies (a) and (b), then
  $N_1(s)=TN(s)$ where $T$ is an $r\times r$ constant unitary matrix.
\item
If $X(s)$ is analytic on the finite
$i\omega$ axis, then $N(s)$ is analytic in a right half-plane
$\mathrm{Re}s>-\tau$, $\exists\tau>0$. If in addition, the normal rank of
$X(s)$ is invariant on the finite $i\omega$ axis, then  $N^{-1}(s)$ is also
analytic in a right half plane $\mathrm{Re}s>-\tau_1$, $\exists\tau_1>0$.    
\item
By applying claims (a)-(d) to $X(s)^T$, one can obtain the
factorization $ X(s)=M(s)M(s)^H$, where the spectral factor $M(s)$ has the
dimension $n\times r$ and has the same analyticity properties as $N(s)$.   
\end{enumerate}
\end{lemma}

Consider a proper rational transfer function $H_{11}(s)$ with the properties
\begin{enumerate}[(H1):]
\item
$H_{11}(s)$ has poles in the open left half-plane of the complex plane, and
is analytic in a right half-plane $\mathrm{Re}s>-\tau$ ($\exists \tau>0$);
\item 
$H_{11}(i\omega)H_{11}(i\omega)^\dagger \le I_n$ $\forall\omega\in\bar{\mathbf{R}}$; and
\item
The normal rank of the following matrices does not change on the finite
imaginary axis $i\omega$: 
\begin{eqnarray}
  \label{eq:26}
  X_1(s)&=&I_n-H_{11}(s)H_{11}(s)^H, \nonumber \\ 
  X_2(s)&=&I_n-H_{11}(s)^HH_{11}(s).
\end{eqnarray}
\end{enumerate}
The transfer functions $X_1(s)$ and $X_2(s)$ defined in~(\ref{eq:26}) are
para-Hermitian, and according to (H2), $X_1(i\omega)$ and 
$X_2(i\omega)$ are positive semidefinite. Therefore, according to
Lemma~\ref{Youla.T2} these matrices admit
spectral factorizations. Let $H_{12}(s)$, $\tilde H_{21}(s)$
  denote spectral factors of $X_1(s)$, $X_2(s)$ such that 
\begin{eqnarray}
X_1(s)=H_{12}(s)H_{12}(s)^H, \ \ 
X_2(s)=\tilde H_{21}(s)^H  \tilde H_{21}(s). \quad \label{eq:85}
\end{eqnarray}
Also, let $\tilde H_{21}^{-1}(s)$ denote the right inverse of $\tilde H_{21}(s)$, 
$ 
\tilde H_{21}(s) \tilde H_{21}^{-1}(s)=I_r,
$ 
where $r$ is the normal rank of $X_2(s)$. 

\begin{theorem}
  \label{two-step}
Given a proper rational transfer function $H_{11}(s)$ which satisfies conditions (H1)--(H3),
let $H_{12}(s)$ and $\tilde H_{21}(s)$ be the spectral
factors from~(\ref{eq:85}). Define
\begin{eqnarray}
  \label{eq:81}
&&H_{21}(s)=U(s)\tilde H_{21}(s), \nonumber \\
&&H_{22}(s)=-U(s)(\tilde H_{21}^{-1}(s))^H H_{11}(s)^H
   H_{12}(s), \quad
\end{eqnarray}
where $U(s)$ is a stable causal paraunitary $r\times r$ transfer function
matrix, chosen to cancel unstable poles of  
$(\tilde H_{21}^{-1}(s))^H H_{11}(s)^H H_{12}(s)$; cf.~\cite{Shaked-1990}. 
The corresponding $(n+r)\times (n+r)$ transfer function 
$H(s)$ in~(\ref{eq:98a}) is stable, causal and
satisfies~(\ref{eq:60}). 
\end{theorem}

\emph{Proof: }
For simplicity of notation, we drop the argument $s$ of the transfer functions.

Since $H_{11}$ is a proper rational stable transfer function, it is causal
according to the Paley-Wiener Theorem~\cite{Yosida}. 
Furthermore since $H_{11}$ is analytic in a
right half-plane $\mathrm{Re}s>-\tau$ ($\exists \tau>0$), 
it is 
analytic on the imaginary axis. Together with the condition that the normal
rank of $H_{11}$ does not change along the imaginary axis, this guarantees
that the spectral factors $H_{12}$, $\tilde H_{21}$ and the right inverse
$\tilde H_{21}^{-1}$ are analytic in a right half-plane
$\mathrm{Re}s>-\tau$, $\exists \tau>0$; see claim (d) of
Lemma~\ref{Youla.T2}. These 
properties ensure that the rational transfer functions $H_{12}$, $\tilde H_{21}$ and
$\tilde H_{21}^{-1}$ are stable and causal; the latter conclusion follows
from the Paley-Wiener Theorem.
Stability and causality of  $H_{21}$, $H_{22}$ now follow from their
definitions expressed in terms of stable causal $H_{11}$, $H_{12}$, $\tilde
H_{21}$ and $\tilde H_{21}^{-1}$. 

We now show that $H$ is paraunitary. The identity~(\ref{eq:9p})
follows directly from the first identity in~(\ref{eq:85}).  Also,
using~(\ref{eq:85}), the identity~(\ref{eq:10p}) can be
verified:  
\begin{eqnarray}
  \lefteqn{H_{11}H_{21}^H+H_{12}H_{22}^H} &&
  \nonumber \\
&=& (H_{11}\tilde H_{21}^H - H_{12} H_{12}^H H_{11}(s)
\tilde H_{21}^{-1})U^H \nonumber \\
&=& (H_{11}\tilde H_{21}^H - H_{11}(I-
    H_{11}^H H_{11}) \tilde H_{21}^{-1})U^H \nonumber \\
&=& (H_{11}\tilde H_{21}^H - H_{11}
    \tilde H_{21}^H \tilde H_{21} \tilde H_{21}^{-1})U^H =0.
  \label{eq:70}
\end{eqnarray}
Furthermore,~(\ref{eq:11p}) also holds:
\begin{eqnarray}
  \lefteqn{H_{21}H_{21}^H+H_{22}H_{22}^H} &&
  \nonumber \\
&=& U (\tilde H_{21}\tilde H_{21}^H + (\tilde
    H_{21}^{-1})^HH_{11}^H H_{12}
    H_{12}^H H_{11}\tilde H_{21}^{-1})U^H \nonumber \\
&=& U (\tilde H_{21}\tilde H_{21}^H \nonumber \\
&& + (\tilde
    H_{21}^{-1})^H(H_{11}^H H_{11}
    -H_{11}^H H_{11}H_{11}^H H_{11})\tilde
    H_{21}^{-1})U^H \nonumber \\ 
&=& U (\tilde H_{21}\tilde H_{21}^H + (\tilde
    H_{21}^{-1})^H(I-\tilde H_{21}^H \tilde H_{21} \nonumber \\
&& -(I-\tilde H_{21}^H \tilde H_{21})(I-\tilde H_{21}^H \tilde
   H_{21}))\tilde H_{21}^{-1})U^H \nonumber \\ 
&=& UU^H =I.
 \label{eq:82}  
\end{eqnarray}
\hfill$\Box$

\begin{remark}\label{H1-H3}
We note that condition
(H2) is necessary for $H(s)$ to satisfy~(\ref{eq:60}). Also,
(H1) and (H2) together ensure that $H_{11}(s)$ is stable and causal which is
necessary for $H(s)$ to have these properties. Condition (H3) is a
technical condition; it ensures that the spectral factor $\tilde H_{21}(s)$ of
$X_2(s)$ is stable and causal and has a stable causal inverse. 
\hfill$\Box$
\end{remark}

\begin{remark}\label{r=0}
Theorem~\ref{two-step} shows that the number of noise channels $z$, $z^\#$
necessary to ensure that the equalizing filter is physically
realizable is determined by the normal rank of
$I_n-H_{11}(s)^HH_{11}(s)$. In particular, when $H_{11}(s)^HH_{11}(s)=I$,
the transfer function $H_{11}(s)$ is physically realizable, and 
additional noise channels are not required.  
\hfill$\Box$
\end{remark}

\subsection{The auxiliary optimization problem}

In the remainder of the paper, the two-step procedure described in the
previous section will pave the way to developing optimization approaches to
solving the quantum equalizer design problem. 
For a constant $\gamma>0$, define the feasible set $\mathcal{H}_{11,\gamma}$ 
consisting of proper rational $n\times n$ transfer function
matrices $H_{11}(s)$, which satisfy conditions~(H1),~(H2) 
and~(\ref{eq:6'.sub}). For convenience, we summarize the two latter
conditions as 
\begin{eqnarray}
  \label{eq:17}
  \label{eq:76}
&&  P_e(i\omega,H_{11})< \gamma^2I_n, \\
&&  H_{11}(i\omega)H_{11}(i\omega)^\dagger \le I_n \quad
   \forall\omega\in\bar{\mathbf{R}}. 
  \label{eq:19}
\end{eqnarray}
In~(\ref{eq:17}), we slightly abuse the notation and write
$P_e(i\omega,H_{11})$ for the expression on the
right hand side of~(\ref{eq:121p}), to emphasize that the independent
variable of this function is $H_{11}(s)$.
Note that feasible $H_{11}$ are elements of the
Hardy space $H_\infty$ and ~(\ref{eq:19}) implies $\|H_{11}\|_\infty\le 1$. 

Theorem~\ref{two-step} leads us to replace
Problem~\ref{P1} with the following auxiliary optimization problem.

\begin{problemprime}\label{P1a}
The auxiliary guaranteed cost problem is to obtain, for a given $\gamma>0$,
a feasible $H_{11}(s)\in \mathcal{H}_{11,\gamma}$. The corresponding
auxiliary optimal filtering problem is to determine an optimal level of
guaranteed performance    
\begin{eqnarray}
  \label{eq:6''}
\gamma_\circ'=\inf\{\gamma>0\colon \mathcal{H}_{11,\gamma}\neq\emptyset\}.
\end{eqnarray}
\end{problemprime}

Formally, one must distinguish between $\gamma_\circ$ defined
  in~(\ref{eq:6'}) and $\gamma_\circ'$ defined in
  Problem~\ref{P1a}. Solutions of the latter problem are not guaranteed to
  satisfy condition (H3) of Theorem~\ref{two-step}, therefore
  $\gamma_\circ'\le \gamma_\circ$. Nevertheless, the 
  remainder of the paper focuses on the auxiliary
  Problem~\ref{P1a}. We will observe later in
  Section~\ref{examples} that the (sub)optimal transfer functions $H_{11}(s)\in
  \mathcal{H}_{11,\gamma}$ obtained in the examples considered in that section
   also satisfy (H3). Therefore, the gap between $\gamma_\circ'$ and
   $\gamma_\circ$ vanishes in those examples. 

Note that when $\gamma^2\ge \boldsymbol{\sigma}(\Sigma_u^T+2I)$
(equivalently, $\gamma^2 I\ge \Sigma_u^T+2I$), the auxiliary guaranteed
cost problem has a trivial solution since for such $\gamma$, the set
$\mathcal{H}_{11,\gamma}$ contains $H_{11}(s)=0$. In this case,    
a trivial suboptimal filter in $\mathcal{H}_\gamma$ can be
  readily constructed using Theorem~1, e.g., 
$H(s)=
\left[
  \begin{array}{cc}
    0 & I \\ I & 0
  \end{array}
\right]$. 
Therefore, the standing assumption in the remainder of the paper is
that 
\begin{equation}
  \label{eq:53}
  \gamma^2< \boldsymbol{\sigma}(\Sigma_u^T+2I).
\end{equation}

\subsection{Relation to the classical $H_\infty$-like
  equalization}\label{S-proc} 

The constraint~(\ref{eq:19}) reflects the distinction between the
coherent equalization problem and its classical $H_\infty$-like
counterpart. The latter involves optimization of the bound on the PSD
matrix $P_e$ but does not include condition~(\ref{eq:19}):   
\begin{eqnarray}
  \label{eq:71.nc}
  && \gamma_*=\inf \gamma, \\
  && \mbox{subject to } \gamma>0,\quad P_{e}(i\omega,H_{11})< \gamma^2 I_n  \quad \forall \omega
   \in\bar{\mathbf{R}}. \nonumber
\end{eqnarray} 
Clearly, for every $\gamma>0$, the set $\mathcal{H}_{11,\gamma}$ of feasible
optimizers $H_{11}$ of Problem~\ref{P1a} is a subset of the feasible set of
the problem~(\ref{eq:71.nc}). On the other hand, later in
the paper we will encounter a situation in which a suboptimal filter
of  problem~(\ref{eq:71.nc}) also satisfies~(\ref{eq:19}). 
Via the S-procedure, such situation can be
related to feasibility of a semidefinite program.  

\begin{theorem}\label{SDP.primal.LMI}
Suppose $\gamma\ge \gamma_*$. If there exists $\theta >0$ such that
$\forall\omega\in\bar{\mathbf{R}}$ 
\begin{equation}
\label{eq:21}
        \theta \left[
       \begin{array}{cc}
\Psi(i\omega) &  -G_{11}(i\omega)(I_n+\Sigma_u^T)\\
-(I_n+\Sigma_u^T)G_{11}(i\omega)^\dagger &~~ \Sigma_u^T+(2-\gamma^2)I_n   
       \end{array}       \right]-J\ge 0, 
\end{equation}
then the feasible set of problem~(\ref{eq:71.nc}) is equal to
the feasible set of Problem~\ref{P1a} $\mathcal{H}_{11,\gamma}$.
\end{theorem}

\emph{Proof: }
After pre- and
postmultiplying~(\ref{eq:21}) by $[H_{11}(i\omega)~I_n]$ and 
$\left[
  \begin{array}{c}
    H_{11}(i\omega)^\dagger \\ I_n
  \end{array}
\right]$, respectively, (\ref{eq:21}) becomes
\[
\theta
(P_e(i\omega,H_{11})-\gamma^2I_n)-(H_{11}(i\omega)H_{11}(i\omega)^\dagger-I_n)\ge 0.
\]   
Now let $H_{11}(s)$ be a feasible transfer function of problem~(\ref{eq:71.nc}),
such that $P_e(i\omega,H_{11})<\gamma^2I$ $\forall\omega\in\bar{\mathbf{R}}$. Then
\[
H_{11}(i\omega)H_{11}(i\omega)^\dagger\le I_n+\theta(P_e(i\omega,H_{11})-\gamma^2I_n)\le I_n. 
\]
That is, $H_{11}(s)\in \mathcal{H}_{11,\gamma}$. This shows
  that under the conditions of the theorem, the feasible set of
  problem~(\ref{eq:71.nc}) is a subset of 
  $\mathcal{H}_{11,\gamma}$. 
Thus, the claim of the
theorem follows, due to the previous observation that the
  converse inclusion also holds.  
\hfill$\Box$

From Theorem~\ref{SDP.primal.LMI}, it follows that under
condition~(\ref{eq:21}), the constraint~(\ref{eq:19}) of
problem~(\ref{eq:6''}) is inactive and any suboptimal filter of
problem~(\ref{eq:71.nc}) is also a guaranteed cost filter for
Problem~\ref{P1a}.

\section{Parameterization of suboptimal causal physically realizable
  filters}\label{feasible} 

In this section, we will derive a parametric representation of the set
$\mathcal{H}_{11,\gamma}$ of feasible
optimizers of Problem~\ref{P1a}, 
given a $\gamma>\gamma_\circ'$. 

The problem of characterizing all
causal rational proper transfer functions which
satisfy~(\ref{eq:76}),~(\ref{eq:19}) is similar to the problem of 
describing suboptimal $H_\infty$ filters for a linear uncertain system,
with the additional constraint  that $\|H_{11}\|_\infty\le 1$. 
We apply the technique of $J$-spectral
factorization~\cite{GGLD-1990,IO-1996} to solve this problem under the following technical assumption. 

\begin{assumption}\label{A2}
  The matrix $\Psi(s)$ in~(\ref{eq:47}) has full normal rank. 
\end{assumption}

Next, we note that $\Psi(s)$ and its transpose $\Psi(s)^T$ are proper rational
para-Hermitian matrices, 
$\Psi(s)^H=\Psi(s)$, $\Psi(-s^*)^\#=\Psi(s)^T$.    
Therefore, according to Lemma~\ref{Youla.T2}, applied to $\Psi(s)^T$, there
exists a rational matrix $M(s)$ such that 
\begin{equation}
  \label{eq:61}
  \Psi(s)=M(s)M(s)^H.
\end{equation}
Under Assumption~\ref{A2}, the matrix $M$ is a square $n\times n$
matrix, and its left inverse $M^{-1}(s)$ is the same as its right inverse. 
Since the matrix $\breve{A}$ of the channel system is assumed to be stable
(Assumption~\ref{A1}), 
$G_{11}(s)$, $G_{12}(s)$ are analytic on the imaginary axis. Therefore
$\Psi(s)$ is also analytic on the imaginary axis and $\Psi(i\omega)\ge 0$ for
all $\omega$. Then according to Lemma~\ref{Youla.T2},  
$M(s)$ and $M^{-1}(s)$ are analytic in a right half-plane $\mathrm{Re}s>-\tau$,
$\exists\tau>0$. These observations allow us to express the expression for 
$P_e(s,H_{11})-\gamma^2I$ as 
\begin{eqnarray}
  \label{eq:57}
&& P_e(s,H_{11})-\gamma^2I_{2n}=
     \left[
     \begin{array}{cc}
      Y(s) & I_n
     \end{array}
          \right]\Phi(s)      \left[
     \begin{array}{c}
      Y(s)^H 
\\ I_n
     \end{array}
  \right], \quad
\end{eqnarray}
where $Y(s) = H_{11}(s)M(s)$, and
\begin{eqnarray}
&& \Phi(s) =
       \left[
       \begin{array}{cc}
I_n & Q(s) \\
Q(s)^H 
         &~~ \Sigma_u^T+(2-\gamma^2)I_n   
       \end{array}       \right],  \nonumber \\
&& Q(s) \triangleq -M^{-1}(s)G_{11}(s)(I_n+\Sigma_u^T), \label{eq:75} 
\end{eqnarray}
$Q(s)$ is analytic in a right half-plane $\mathrm{Re}s>-\tau$,
$\exists\tau>0$.

Recall that $(2n)\times (2n)$ rational matrix transfer function $\Phi(s)$
is said to admit a (left-standard) $J$-spectral factorization
if it can be represented as 
\begin{equation}
  \label{eq:68}
  \Phi(s)=\Upsilon(s)J \Upsilon(s)^H, 
\end{equation}
where a $(2n)\times (2n)$ rational transfer matrix $\Upsilon(s)$ has all
its poles in the left half-plane $\mathrm{Re}s<-\tau_2$ ($\exists
\tau_2>0$)\footnote{Normally, $J$-spectral factors $\Upsilon(s)$ are required
  to be analytic and bounded in the right half-plane
  $\mathrm{Re}s>0$~\cite{FD-1987} or have poles in the left half-plane
  $\mathrm{Re}s<0$~\cite{IO-1996,GGLD-1990}. Our somewhat
  stronger requirements are dictated by the requirement that 
  the spectral factor $M(s)$ of $\Psi(s)$ must be invertible on the
  imaginary axis and that $M(s)^{-1}$ must also be analytic in the right half
  plane and $M(i\omega)^{-1}$ must be well defined on the imaginary
  axis. To meet these requirements, we employ
  Lemma~\ref{Youla.T2} which 
  requires that $\Psi(s)$ must be analytic on the imaginary axis.}
~\cite{IO-1996,GGLD-1990}. 

The following theorem adapts Theorem~1 of~\cite{IO-1996} to
  the left-standard factorization setting of this paper.  

\begin{theorem}\label{T2}
Suppose Assumption~\ref{A2} is satisfied and there exists a spectral factor $M(s)$ defined in~(\ref{eq:61})
such that the $(2n)\times (2n)$ transfer matrix $\Phi(s)$
in~(\ref{eq:75}) has a $J$-spectral factorization~(\ref{eq:68}), where 
\begin{equation}
  \label{eq:38}
\Upsilon(s)=
\left[
  \begin{array}{cc}
   \Upsilon_1(s) & \Upsilon_2(s) \\
   \Upsilon_3(s) & \Upsilon_4(s)
  \end{array}
\right],  
\end{equation}
$\Upsilon_{j}(s)$, $j=1,2,3,4$, and also the inverses $\Upsilon(s)^{-1}$,
$\Upsilon_1(s)^{-1}$ are analytic in a right half-plane
$\mathrm{Re}s>-\tau_2$ and have their poles in the half-plane
$\mathrm{Re}s< -\tau_2$ ($\exists \tau_2>0$). Then 
$H_{11}(s)\in \mathcal{H}_{11,\gamma}$ if and only if 
\begin{equation}
     \label{eq:67}
   H_{11}(s)=S_2^{-1}(s)S_1(s)M^{-1}(s),
\end{equation}
where 
\begin{equation}
  \label{eq:77}
  \left[
    \begin{array}{cc}
      S_1(s) & S_2(s)
    \end{array}
  \right] = \left[
    \begin{array}{cc}
      \Theta(s) & I_n
    \end{array}
  \right] \Upsilon(s)^{-1}
\end{equation}
for a rational stable  $n\times n$ transfer function matrix $\Theta(s)\in
H_\infty$ 
analytic in a right half-plane $\mathrm{Re}s>-\tau$ ($\exists \tau>0$),
such that $\|\Theta\|_\infty<1$, and also  
\begin{eqnarray}
  \label{eq:79}
  S_1(i\omega)M(i\omega)^{-1}(M(i\omega)^{-1})^\dagger S_1(i\omega)^\dagger
  && \nonumber \\
\le 
  S_2(i\omega)S_2(i\omega)^\dagger && \quad \forall \omega\in
                                       \bar{\mathbf{R}}.\quad  
\end{eqnarray}
\end{theorem}

\emph{Proof: }
\emph{The ‘only if’ claim}: The statement $H_{11}\in\mathcal{H}_{11,\gamma}$
reads that the transfer function matrix $H_{11}(s)$ is a 
rational transfer function matrix which is stable, analytic in a right 
half-plane $\mathrm{Re}s>-\tau$ ($\exists \tau>0$) and satisfies
conditions~(\ref{eq:76})
and~(\ref{eq:19}). Then the matrix $Y(s)=H_{11}(s)M(s)$ is also stable and
analytic in a right
half-plane $\mathrm{Re}s>-\tau_1$, $\exists\tau_1>0$, and
from~(\ref{eq:76}) we have
\begin{equation}
  \label{eq:80}
\left[
     \begin{array}{cc}
      Y(i\omega) & I_n
     \end{array}
          \right]\Phi(i\omega)      \left[
     \begin{array}{c}
      Y(i\omega)^\dagger \\ I_n
     \end{array}
  \right]<0 \quad \forall \omega\in \bar{\mathbf{R}}.    
\end{equation}
Following the same lines that were used to prove Theorem~1
in~\cite{IO-1996}, one can show that the matrices $\Theta_1(s)$,
$\Theta_2(s)$ defined by the equation 
\[
  \left[
    \begin{array}{cc}
      \Theta_1(s) & \Theta_2(s)
    \end{array}
  \right] = \left[
    \begin{array}{cc}
      Y(s) & I_n
    \end{array}
  \right] \Upsilon(s)
\]
are analytic in a right half-plane $\mathrm{Re}s>-\tau$ ($\exists\tau>0$),
stable and that  
\begin{equation}
  \label{eq:84}
  \Theta_1(i\omega)\Theta_1(i\omega)^\dagger<\Theta_2(i\omega)\Theta_2(i\omega)^\dagger
  \quad  \forall \omega\in \bar{\mathbf{R}}.
\end{equation}
In particular, it follows from~(\ref{eq:84}) that $\Theta_2(s)$ is
invertible on the imaginary axis, and that
$\|\Theta\|_\infty<1$ where 
$\Theta(s)=\Theta_2(s)^{-1}\Theta_1(s)$. 
Furthermore, $\Theta_2^{-1}(s)$ has all its poles in the left
  half-plane~\cite{IO-1996}, thus $\Theta(s)$ is analytic in a right half-plane $\mathrm{Re}s>-\tau$ ($\exists\tau>0$). Also 
\[
  \left[
    \begin{array}{cc}
      \Theta_2(s)^{-1} Y(s)~ & \Theta_2(s)^{-1}
    \end{array}
  \right] = \left[
    \begin{array}{cc}
     \Theta(s) & I
    \end{array}
  \right] \Upsilon^{-1}(s).
\]
Letting $S_1(s)=\Theta_2(s)^{-1} Y(s)=\Theta_2(s)^{-1} H_{11}(s)M(s)$,
$S_2(s)=\Theta_2(s)^{-1}$ yields~(\ref{eq:77}). We also can express
$H_{11}(s)$ as $H_{11}(s)=S_2(s)^{-1}S_1(s)M(s)^{-1}$. This
gives~(\ref{eq:67}). Substituting this expression into~(\ref{eq:19})
results in~(\ref{eq:79}).      

{\emph{The `if' claim:}
This part of the proof also replicates the proof of the
corresponding statement in Theorem~1 of~\cite{IO-1996}. Using the same
reasoning as in that theorem, one can show that $S_2(s)$ in
equation~(\ref{eq:77}) is invertible and that its inverse is stable and
analytic in a right half-plane $\mathrm{Re}s>-\tau$, $\exists\tau>0$. Also,
since 
$\|\Theta\|_\infty<1$,  
then using~(\ref{eq:68}) we obtain   
\begin{eqnarray*}
\lefteqn{  \left[
  \begin{array}{cc}
  S_1(i\omega)  & S_2(i\omega)
  \end{array}
  \right]\Phi(i\omega)\left[
     \begin{array}{c}
      S_1(i\omega)^\dagger  \\ S_2(i\omega)^\dagger
     \end{array}\right]} && \\
&& = \left[
  \begin{array}{cc}
  S_1(i\omega)  & S_2(i\omega)
  \end{array}
  \right]\Upsilon(i\omega)J \Upsilon(i\omega)^\dagger\left[
     \begin{array}{c}
      S_1(i\omega)^\dagger  \\ S_2(i\omega)^\dagger
     \end{array}\right] \\
&& = \left[
  \begin{array}{cc}
  \Theta(i\omega)  & I
  \end{array}
  \right]J \left[
     \begin{array}{c}
      \Theta(i\omega)^\dagger  \\ I
     \end{array}\right] <0 \quad \forall\omega\in\bar{\mathbf{R}}.
\end{eqnarray*}
Therefore, we conclude that $Y(s)=S_2(s)^{-1}S_1(s)$
satisfies~(\ref{eq:80}). Therefore, $H_{11}(s)=Y(s)M^{-1}(s)$
satisfies~(\ref{eq:17}). Also, (\ref{eq:79}) implies that this $H_{11}$
satisfies~(\ref{eq:19}). Furthermore, this transfer 
function matrix $H_{11}(s)$ has the required stability and analyticity
properties to be an element of $\mathcal{H}_{11,\gamma}$. 
\hfill$\Box$

Adding the inequality (\ref{eq:21}) from Theorem~\ref{SDP.primal.LMI} to
the condition of Theorem~\ref{T2} will render the inequality~(\ref{eq:79})
redundant. As a result, we have the following corollary.  

\begin{corollary}\label{cor}
Suppose that the conditions of Theorem~\ref{T2} hold and, in addition, the
inequality (\ref{eq:21}) from Theorem~\ref{SDP.primal.LMI} also holds for
some $\theta> 0$. Then $H_{11}\in \mathcal{H}_{11,\gamma}$ if
  and only if it can be represented in 
the form~(\ref{eq:67}), where $S_1(s)$, $S_2(s)$ are determined as
in~(\ref{eq:77}) using a stable rational transfer function matrix
$\Theta(s)$ analytic in a right half-plane $\mathrm{Re} s>-\tau$
($\exists\tau>0$) and such that $\|\Theta\|_\infty<1$.  
\end{corollary}

The following corollary is concerned with a special case of Theorem~\ref{T2}
where $\Upsilon_4(s)=0$ in $\Upsilon(s)$. The corollary shows
that in this special case, the transfer function $H_{11}(s)$ can be
expressed in the form resembling the celebrated Youla parameterization of
all stabilizing controllers in the classical $H_\infty$ control
problem~\cite{ZDG-1996}. This special case will prove useful in
the examples considered in Section~\ref{examples}. 

\begin{corollary}\label{cor.LFT}
Suppose that the conditions of Theorem~\ref{T2} hold and, in addition, the spectral factor $\Upsilon(s)$ has $\Upsilon_4(s)=0$, and
$\Upsilon_2(s)$, $\Upsilon_3(s)$ are invertible in the right half-plane
$\mathrm{Re}s>-\tau$ ($\exists \tau>0$). Then every feasible $H_{11}(s)$
has the form
\begin{eqnarray}
     \label{eq:67.LFT}
   H_{11}(s)&=&-\Upsilon_3(s)
(I-\Upsilon_1^{-1}(s)\Upsilon_2(s)\Theta(s))^{-1} \nonumber \\
&&\times \Upsilon_1^{-1}(s)M^{-1}(s). 
\end{eqnarray}
where $\Theta(s)$ is a stable rational $n\times n$ transfer function matrix
analytic in a right half-plane $\mathrm{Re}s>-\tau$ ($\exists \tau>0$)
such that $\|\Theta\|_\infty<1$ and
\begin{eqnarray}
  \label{eq:36}
\lefteqn{M(i\omega)^{-1}(M(i\omega)^{-1})^\dagger\le 
(\Upsilon_2(i\omega)\Theta(i\omega)-\Upsilon_1(i\omega))
} &&   \nonumber \\ 
&& \times \Upsilon_3(i\omega)^{-1}
(\Upsilon_3(i\omega)^{-1})^\dagger
(\Upsilon_2(i\omega)\Theta(i\omega)-\Upsilon_1(i\omega))^\dagger \nonumber\\
&& \qquad\qquad \forall \omega\in\bar{\mathbf{R}} . 
\end{eqnarray}
\end{corollary}

\emph{Proof: }
The statement of the corollary 
follows directly from~(\ref{eq:67}) and~(\ref{eq:77}) using
the fact that 
\begin{equation*}
\Upsilon(s)^{-1}=
\left[
  \begin{array}{cc}
   0 & \Upsilon_3^{-1}(s) \\
   \Upsilon_2^{-1}(s) &~~ -\Upsilon_2^{-1}(s) \Upsilon_1(s)\Upsilon_3^{-1}(s) 
  \end{array}
\right].
\end{equation*}
With this expression for $\Upsilon(s)^{-1}$, (\ref{eq:77}) reduces to
\begin{eqnarray}
  \label{eq:45.LFT}
  S_1(s)&=&\Upsilon_2^{-1}(s), \nonumber \\
  S_2(s)&=&-\Upsilon_2^{-1}(s)\Upsilon_1(s)(I-\Upsilon_1^{-1}(s)\Upsilon_2(s)\Theta(s))\Upsilon_3^{-1}(s). \quad\quad
\end{eqnarray}
Substituting these expressions
in~(\ref{eq:67}),~(\ref{eq:79}) leads
to~(\ref{eq:67.LFT}),~(\ref{eq:36}), respectively. 
\hfill$\Box$   

We conclude this section by stressing that combining Theorem~\ref{T2} with
Theorem~\ref{two-step} allows 
  one to obtain a suboptimal coherent equalizing filter $H(s)$ for which
  the power spectral density of the equalization error is guaranteed to
  be bounded from above as in~(\ref{eq:76}). To obtain such a filter,
  $\Theta(s)$ must be chosen to ensure that $H_{11}(s)$ defined
  in~(\ref{eq:67}) also satisfies condition (H3) of
  Theorem~\ref{two-step}. Furthermore, it is also possible to obtain the
  smallest $\gamma^2$ for which conditions of Theorem~\ref{T2} hold. The
  examples in Section~\ref{examples} illustrate these points. The following
  expansion of the $J$-spectral factorization formula~(\ref{eq:68}) will be
  used in these examples 
\begin{eqnarray}
    \label{eq:90}
   && \Upsilon_1(s)\Upsilon_1(s)^H-\Upsilon_2(s)\Upsilon_2(s)^H=I, \nonumber
    \\
   &&
      \Upsilon_3(s)\Upsilon_3(s)^H-\Upsilon_4(s)\Upsilon_4(s)^H=\Sigma_u^T+(2-\gamma^2)I, \nonumber \\
   && \Upsilon_1(s)\Upsilon_3(s)^H-\Upsilon_2(s)\Upsilon_4(s)^H=Q(s).
  \end{eqnarray}

\section{Suboptimal solution via Semidefinite Programming and
  Nevanlinna-Pick interpolation}\label{semidef} 

Theorem~\ref{T2} reduces the coherent passive equalization problem
to finding the smallest constant $\gamma^2$
for which the matrix $\Phi(s)$ admits a $J$-spectral decomposition and for
which a matrix $\Theta(s)$ can be found which satisfies~(\ref{eq:79}). 
In general, finding $J$-spectral factors is known to be a difficult 
problem. Therefore in this section we consider an alternative approach
in which we seek to construct a physically realizable
equalizer which is suboptimal in the sense that it minimizes the
power spectrum density $P_e(i\omega,H_{11})$ at selected frequency
points.  

The proposed approach is based on the observation that
Problem~\ref{P1a} is equivalent to the semidefinite program (SDP)
\begin{eqnarray}
  \label{eq:71.LMI}
  && \inf \gamma^2 \\
  && \mbox{s.t. }     \left[
     \begin{array}{cc}
Z_{11}(i\omega) & H_{11}(i\omega)M(i\omega) \\
M(i\omega)^\dagger H_{11}(i\omega)^\dagger & -I_q 
     \end{array}
     \right]< 0, \label{eq:13} \quad \\
  && \phantom{\mbox{s.t. }} 
     \left[
     \begin{array}{cc}
       I_n & H_{11}(i\omega) \\ H_{11}(i\omega)^\dagger & I_n
     \end{array}
     \right]\ge 0 \quad \forall\omega\in\bar{\mathbf{R}}, \label{eq:14}
\end{eqnarray}
where
\begin{eqnarray*}
Z_{11}(s)&\triangleq&
(2-\gamma^2)I+\Sigma_u^T-H_{11}(s)G_{11}(s)(I+\Sigma_u^T) \\
&& -(I+\Sigma_u^T)G_{11}(s)^H H_{11}(s)^H.  
\end{eqnarray*}
Indeed, (\ref{eq:13}),~(\ref{eq:14}) follow
from~(\ref{eq:17}),~(\ref{eq:19}) using the Schur complement~\cite{HZ-2005}.

The LMI constraints of the problem~(\ref{eq:71.LMI})--(\ref{eq:14}) are
parameterized by the frequency parameter $\omega$. Unless a closed form
solution to this problem can be found, to obtain a numerical solution one
has to resort to a relaxation of the constraints. One such relaxation
involves a grid of frequency points $\omega_l$, $l=1,\ldots,L$:
\begin{eqnarray}
  \label{eq:71.LMI.l}
  && \inf \gamma^2 \\
  && \mbox{s.t. }
     \left[
     \begin{array}{cc}
Z_{11}(i\omega_l) & H_{11,l}M(i\omega_l) \\
M(i\omega_l)^\dagger H_{11,l}^\dagger & -I_q 
     \end{array}
     \right]< 0, \label{eq:13.l} \\
  && \phantom{\mbox{s.t. }} 
     \left[
     \begin{array}{cc}
       I_n & H_{11,l} \\ H_{11,l}^\dagger & I_n
     \end{array}
     \right]\ge 0, \quad l=1, \ldots,L. \label{eq:14.l}
\end{eqnarray}
While this relaxation of the constraints makes the SDP problem more
tractable, it also needs to be complemented with 
interpolation, to obtain a transfer function $H_{11}(s)$ from
which a physically realizable equalizer $\Delta(H(s),0)$ can be obtained. This
requires that the resulting transfer function
matrix $H_{11}(s)$ must satisfy~(\ref{eq:19}).
To accomplish this, we use the Nevanlinna-Pick interpolation of the solution of
the relaxed problem~(\ref{eq:71.LMI.l})--(\ref{eq:14.l}).  

Recall the formulation of the matrix Nevanlinna-Pick interpolation
problem~\cite{BGR-2013,DGK-1979,Kovalishina-1984}. Our
  formulation follows~\cite{Kovalishina-1984}, which gives a solution
  of the
  kind of the 
Nevanlinna-Pick problem which is most convenient for
application to the 
problem considered in this section. Given a set of
distinct points $\{s_l, l=1, \ldots, L\}$ located in the open 
  right half-plane $\mathrm{Re}s>0$, and a collection of $n\times n$
matrices $\{\mathbf{X}_l, 
l=1,\ldots,L\}$~\footnote{In the most general setting, the index set on
  which $l$ varies is 
  not necessarily finite, nor even countable~\cite{DGK-1979}.}, the
matrix Nevanlinna-Pick interpolation consists in finding a rational
$n\times n$  matrix-valued function 
$\mathbf{X}(s)$ which is analytic in the open
  right half-plane $\mathrm{Re}s>0$, satisfies
$\mathbf{X}(s_l)=\mathbf{X}_l$ and such that  
$\|\mathbf{X}(s)\|\le 1$ for $\mathrm{Re}s>0$. Here
$\|\mathbf{X}(s)\|$ is the 
spectral norm of the matrix $\mathbf{X}(s)$, i.e., the largest singular value of
$\mathbf{X}(s)$\footnote{In~\cite{Kovalishina-1984},
    this requirement is expressed as $I-\mathbf{X}(s)^\dagger \mathbf{X}(s)\ge 0$.}.   

We now summarize the algorithm for finding a suboptimal solution to
Problem~\ref{P1a} which is based on the results
  of~\cite{Kovalishina-1984}. Given a collection of frequencies
$\omega_l$, $l=1, \ldots,L$, let $\tilde \gamma>0$ and $H_{11,l}$, $l=1,
\ldots,L$, be a (suboptimal) solution of the LMI optimization
problem~(\ref{eq:71.LMI.l})--(\ref{eq:14.l}).  
Let $\tau$ be a sufficiently small positive constant, and define
$s_l=i\omega_l+\tau$. Furthermore, suppose that 
the $nL\times nL$ block-Pick matrix $\mathbf{P}$ consisting of the blocks
\begin{equation}
\label{eq:48}
\mathbf{P}_{l,k}= 
\frac{I-H_{11,l}H_{11,k}^\dagger }{s_l+s_k^*}, \quad l,k=1,\ldots L,
\end{equation}
is positive definite. The matrix version of the 
Nevanlinna~criterion~\cite{Kovalishina-1984} states that this 
is necessary and sufficient  
for the existence of a rational matrix $\hat H_{11}(s)$
which is analytic in $\{s: \mathrm{Re}s>0\}$, satisfies
  $\|\hat H_{11}(s)\|\le 1$ in that domain and such that 
$  \hat H_{11}(s_l)=H_{11,l}$.

A rational interpolant $\hat H_{11}(s)$ can be obtained using the
matrix extension of the Nevanlinna
algorithm~\cite{Kovalishina-1984}; also see
\cite{DGK-1979,GK-1989}. Namely, $\hat H_{11}(s)$ is
  representable as a linear fractional transformation of an arbitrary
  rational stable transfer function
  $\Theta(s)$ which is analytic in $\mathrm{Re}s>0$ and satisfies
  $\|\Theta\|_\infty<1$: 
  \begin{eqnarray}
    \label{eq:58}
   \hat H_{11}(s)&=&(W_{11}(s)\Theta(s)+W_{12}(s)) \nonumber \\
&& \times (W_{21}(s)\Theta(s)+W_{22}(s))^{-1}. 
  \end{eqnarray}
The coefficient matrix of this transformation
\begin{equation}
  \label{eq:83}
  W(s)=
  \left[\begin{array}{cc}
    W_{11}(s) & W_{12}(s) \\
    W_{21}(s) & W_{22}(s)
        \end{array} 
      \right]
\end{equation}
is constructed from the matrix $\mathbf{P}>0$:
\begin{eqnarray}
  \label{eq:101}
  W(s)&=&I-
  \left[
  \begin{array}{ccc}
   \frac{I}{s+s_1^*} & \ldots & \frac{I}{s+s_L^*}\\
    \frac{H_{11,1}^*}{s+s_1^*} & \ldots &\frac{H_{11,L}^*}{s+s_L^*} 
  \end{array}  
  \right] \mathbf{P}^{-1}  \left[\begin{array}{cc}
    I~& -H_{11,1} \\
    \vdots & \vdots \\
    I~ & -H_{11,L} 
  \end{array}  
  \right].\quad  
\end{eqnarray} 

It remains to obtain $H_{11}(s)$. Let       
\begin{equation}
  \label{eq:52}
H_{11}(s)=\hat H_{11}(s+\tau).
\end{equation}
From the properties of $\hat H_{11}(s)$,
it follows that $H_{11}(s)$ is analytic in the half-plane
$\mathrm{Re}s>-\tau$ and $\|H_{11}(s)\|\le 1$ for all $s$ such that
$\mathrm{Re}s> -\tau$. Consequently, $\|H_{11}(i\omega)\|\le 1$ for all
$\omega\in\mathbf{R}$ which
implies~(\ref{eq:19}). Finally, it follows from the definition of
$H_{11}(s)$ and~(\ref{eq:13.l}) that
\begin{equation}
  \label{eq:71}
P_e(i\omega_l, H_{11})< \tilde\gamma^2, \quad l=1, \ldots, L.  
\end{equation}
That is, the constructed transfer function $H_{11}(s)$ is 
suboptimal in the sense that it minimizes the power spectrum density
$P_e(i\omega,H_{11})$ at the selected frequency grid points.    

\begin{remark}\label{rem-posdef}
The requirement of the algorithm that the
block-Pick matrix $\mathbf{P}$ must be positive definite is not
restrictive. It can be satisfied by further restricting~(\ref{eq:14.l}) to
be a strict inequality, and then choosing a sufficiently small $\tau>0$.
Indeed, when  $H_{11,l}^\dagger
H_{11,l}< I$, it follows from~(\ref{eq:14.l}) that the
$(l,l)$-block~(\ref{eq:48}) is positive definite. Its eigenvalues 
can be made arbitrarily large by selecting $\tau>0$ to be sufficiently
close to 0, since the denominator is equal to $2\tau$ and
  vanishes as $\tau\to 0$. The off-diagonal blocks
remain bounded as $\tau\to 0$, making the block-Pick matrix $\mathbf{P}$
block-diagonally dominant~\cite{FV-1962}, with positive definite blocks on the
diagonal. Then if $\tau>0$ is sufficiently small, $\mathbf{P}$ is positive
definite~\cite{ZLHX-2010}.  
\end{remark}

\section{Examples}\label{examples}

\subsection{Coherent equalization of a static two-input two-output system}
\label{ex1.revisited}

Consider a two-input two-output system which mixes a single mode input
field $u$ with a single mode environment field $w$; its outputs and inputs are
related via a static unitary transformation: 
\begin{eqnarray}
  \label{eq:1.bs}
  \left[
    \begin{array}{c}
      y \\ d
    \end{array}
  \right]=  G \left[
    \begin{array}{c}
      u \\  w
    \end{array}
  \right], \quad G=\left[
    \begin{array}{cc}k & m \\
        -e^{i\phi}m^* & e^{i\phi}k^*
    \end{array}
  \right]; \quad
\end{eqnarray}
$k,m$ are complex numbers, $|k|^2+|m|^2=1$, and $\phi$ is a real
number. One example of such system is an optical beamsplitter. 
Beamsplitters play an important role in many
quantum optics applications such as interferometry, holography, laser
systems, etc. The device has two inputs. The input $u$ represents the
signal we would like to split, and the second input $w$ represents the
thermal noise input from the environment. These input fields are related to the
output via a unitary  transformation~(\ref{eq:1.bs}). 

Since both input
  fields are scalar, $u$ and $w$ are scalar operators, and 
  $\Sigma_u$ and $\Sigma_w$ are 
  real constants. To emphasize this, we write $\Sigma_u=\sigma_u^2$,
  $\Sigma_w=\sigma_w^2$. We now illustrate application of Theorem~\ref{T2}
  in this example.  

Using the above notation, $\Psi(s)$ defined in (\ref{eq:47}) is a constant
expressed as $\Psi(s)=\psi=|k|^2\sigma_u^2+|m|^2\sigma_w^2$. The
expression for the power spectrum density matrix~(\ref{eq:121p}) becomes 
\begin{eqnarray}
\label{eq:121ps}
P_{e}(i\omega,H_{11})&=& \psi |H_{11}(i\omega)|^2  \nonumber \\
&-&2(1+\sigma_u^2)\mathrm{Re}[kH_{11}(i\omega)]+ \sigma_u^2+2.
\end{eqnarray}
Assumption~\ref{A2} is satisfied when at least one of the addends in the
expression for $\psi$ is
positive. We suppose in this example that this requirement is
satisfied. Then one can select $M(s)=\psi^{1/2}e^{i\varphi}$, where
  $\psi^{1/2}$ is the real positive root, and $\varphi$ is an arbitrary
  real constant.  Also, $Q(s)$ defined in~(\ref{eq:75}) is constant, 
$Q(s)=-\frac{k(1+\sigma_u^2)e^{-i\varphi}}{\psi^{1/2}}$.

Note that condition~(\ref{eq:53}) reduces to
$\gamma^2<\sigma_u^2+2$. Therefore, we assume that this condition holds.
Also, suppose that
\begin{equation}
  \label{eq:105}
  \gamma^2>
  \begin{cases}
    \sigma_u^2+2-2|k|(1+\sigma_u^2)+\psi, & \text{if $\psi\le
      |k|(1+\sigma_u^2)$}, \\
    \sigma_u^2+2-\frac{|k|^2(1+\sigma_u^2)^2}{\psi}, & \text{if $\psi>
      |k|(1+\sigma_u^2)$}.
  \end{cases}
\end{equation}
Under this condition,
$\frac{|k|^2(1+\sigma_u^2)^2}{\psi(\sigma_u^2+2-\gamma^2)}-1\ge
0$. Therefore, we let 
\begin{eqnarray}
  \label{eq:106}
  \Upsilon_1&=&
  -\frac{k(1+\sigma_u^2)e^{-i\varphi}}{\sqrt{\psi(\sigma_u^2+2-\gamma^2)}}, \quad
  \Upsilon_3=\sqrt{\sigma_u^2+2-\gamma^2},
                \nonumber \\
  \Upsilon_2&=&e^{-i\varphi}\sqrt{\frac{|k|^2(1+\sigma_u^2)^2}{\psi(\sigma_u^2+2-\gamma^2)}-1},
\end{eqnarray}
where the square roots are chosen to be real and positive.

\begin{proposition}\label{Prop1.J}
Suppose $\gamma^2<\sigma_u^2+2$ satisfies condition~(\ref{eq:105}).
Then $H_{11}(s)\in\mathcal{H}_{11,\gamma}$ if and only if it
  can be represented as 
\begin{eqnarray}
  \label{eq:46}
\lefteqn{H_{11}(s)} && \nonumber \\
&& =\frac{\sigma_u^2+2-\gamma^2}{k(1+\sigma_u^2)+\Theta(s)
    \sqrt{|k|^2(1+\sigma_u^2)^2-\psi(\sigma_u^2+2-\gamma^2)}}, \nonumber
  \\
\end{eqnarray}
where $\Theta(s)$ is a stable rational transfer function analytic in the
closed right half-plane, which satisfies $\|\Theta\|_\infty<1$ and the
frequency domain condition
\begin{eqnarray}
  \label{eq:9}
\lefteqn{\sigma_u^2+2-\gamma^2} && \nonumber \\
 &&\le
   |k(1+\sigma_u^2)+\Theta(i\omega)
    \sqrt{|k|^2(1+\sigma_u^2)^2-\psi(\sigma_u^2+2-\gamma^2)}|.\nonumber \\
\end{eqnarray}
One choice of $\Theta$ which satisfies these requirements is
\begin{eqnarray}
  \label{eq:108}
  \Theta&=&
  \begin{cases}
    \epsilon\frac{k}{|k|}& \text{if $\psi\le |k|(1+\sigma_u^2)$}, \\
    0, & \text{if $\psi> |k|(1+\sigma_u^2)$}, 
  \end{cases}
\end{eqnarray}
where $0<\epsilon<1$ must be chosen to be sufficiently close to 1. 
\end{proposition}

\emph{Proof: }
The direct calculation shows that the matrix $\Upsilon$ defined
in~(\ref{eq:38}) with $\Upsilon_1$,
$\Upsilon_2$, $\Upsilon_3$ defined in~(\ref{eq:106}) and $\Upsilon_4=0$
is the $J$-spectral factor of 
\[
\Phi=
\left[\begin{array}{cc}
  1 & -\frac{k(1+\sigma_u^2)e^{-i\varphi}}{\psi^{1/2}} \\
-\frac{k^*(1+\sigma_u^2)e^{i\varphi}}{\psi^{1/2}} &~~ \sigma_u^2+2-\gamma^2
      \end{array}\right].
\]
Thus, the conditions of Theorem~\ref{T2} are
satisfied. Therefore, 
$H_{11}(s)\in \mathcal{H}_{11,\gamma}$ if and only if it can be expressed 
by equation~(\ref{eq:67.LFT}) in which $\Theta(s)$ must satisfy
$\|\Theta\|_\infty<1$ and~(\ref{eq:36}); see
Corollary~\ref{cor.LFT}. Substituting the values $\Upsilon_1$,
$\Upsilon_2$, $\Upsilon_3$ defined in~(\ref{eq:106}) into~(\ref{eq:67.LFT}),~(\ref{eq:36})
yields~(\ref{eq:46}),~(\ref{eq:9}).  
This proves the first part of the proposition.

Next, consider $\Theta$ suggested in~(\ref{eq:108}). It is obvious that
$\|\Theta\|_\infty<1$. Let us show that this $\Theta$
satisfies~(\ref{eq:9}) as well.
First, consider the case where $\psi\le |k|(1+\sigma_u^2)$. 
When $\gamma^2\ge \sigma_u^2+2-|k|(1+\sigma_u^2)$ it holds that
\begin{eqnarray*}
\lefteqn{\sigma_u^2+2-\gamma^2 \le  |k|(1+\sigma_u^2)} && \\
 &&\le
   |k|(1+\sigma_u^2)+\epsilon
    \sqrt{|k|^2(1+\sigma_u^2)^2-\psi(\sigma_u^2+2-\gamma^2)} \\
 &&=
   \left|k(1+\sigma_u^2)+\epsilon\frac{k}{|k|}
    \sqrt{|k|^2(1+\sigma_u^2)^2-\psi(\sigma_u^2+2-\gamma^2)}\right|.
\end{eqnarray*}
for any $\epsilon\in(0,1)$. Therefore~(\ref{eq:9}) holds in this case. 

When $\sigma_u^2+2-2|k|(1+\sigma_u^2)+\psi < \gamma^2\le
\sigma_u^2+2-|k|(1+\sigma_u^2)$, the left hand-side of this inequality
implies that 
\begin{eqnarray*}
\lefteqn{ 0\le \sigma_u^2+2-\gamma^2-|k|(1+\sigma_u^2)} && \\
&&< \sqrt{|k|^2(1+\sigma_u^2)^2-\psi(\sigma_u^2+2-\gamma^2)}.
\end{eqnarray*}
Since the rightmost inequality is strict, one can choose $\epsilon\in(0,1)$
which is sufficiently close to $1$ and still ensures that
\begin{eqnarray*}
\lefteqn{ 0\le \sigma_u^2+2-\gamma^2-|k|(1+\sigma_u^2)} && \\
&&\le \epsilon \sqrt{|k|^2(1+\sigma_u^2)^2-\psi(\sigma_u^2+2-\gamma^2)}.
\end{eqnarray*}
Thus, we again obtain that~(\ref{eq:9}) holds. 

Now consider the case where $\psi>|k|(1+\sigma_u^2)$. Since in this case we
assume that 
\begin{eqnarray*}
\gamma^2 &>& \sigma_u^2+2-\frac{|k|^2(1+\sigma_u^2)^2}{\psi} \ge \sigma_u^2+2-|k|(1+\sigma_u^2)
\end{eqnarray*}
then $\sigma_u^2+2-\gamma^2\le |k|(1+\sigma_u^2)$. Thus, (\ref{eq:9})
holds with $\Theta=0$.
\hfill$\Box$

Proposition~\ref{Prop1.J} shows that for every $\gamma$ such that
$\gamma^2<\sigma_u^2+2$ and which satisfies the condition~(\ref{eq:105}),
there exists a transfer function $H_{11}(s)$ with the desired 
properties~(\ref{eq:17}),~(\ref{eq:19}). When $\Theta$ is chosen according
to~(\ref{eq:108}), the corresponding $H_{11}$ also satisfies the conditions
of Theorem~\ref{two-step} including condition (H3). Hence,  
for each such $\gamma$, an equalizer
$\Xi=\Delta(H,0)\in\mathcal{H}_p$ can be 
constructed which guarantees that the corresponding error power
  spectrum density $P_e$ does not exceed $\gamma^2$. This leads
to the upper bound on the optimal equalization
performance achievable by coherent passive filters for this system 
\begin{equation}
  \label{eq:23}
(\gamma_\circ')^2\le \gamma_\circ^2=\inf_{H\in\mathcal{H}_p}P_e\le \begin{cases}
\psi-2(1+\sigma_u^2)|k|+(2+\sigma_u^2), \\
& \hspace{-3cm} \text{if $\psi\le (1+\sigma_u^2)|k|$}; \\
(2+\sigma_u^2)-\frac{(1+\sigma_u^2)^2|k|^2}{\psi}, \\ 
& \hspace{-3cm} \text{if $\psi> (1+\sigma_u^2)|k|$}.
  \end{cases}
\end{equation}
Indeed, from Proposition~\ref{Prop1.J} and the remark
  following the statement of Problem~\ref{P1a} we have 
\begin{eqnarray*}
  (\gamma_\circ')^2\le \gamma_\circ^2=\inf_{H\in\mathcal{H}_p}P_e\le P_e(H_{11})<\gamma^2.
\end{eqnarray*}
Taking infimum over $\gamma$ subject to~(\ref{eq:105}) yields~(\ref{eq:23}).  

It turns out that this upper bound is in fact
tight; i.e., the inequalities in~(\ref{eq:23}) are in fact the equalities. The
matrix $H_{11}$ that gives rise to the filter which attains the infimum
in~(\ref{eq:23}) can be obtained as a limit of the
matrix~(\ref{eq:46}),~(\ref{eq:108}) as $\gamma\to \gamma_\circ$:
\begin{equation}
  \label{eq:5}
  H_{11}=\begin{cases}
\frac{k^*}{|k|}, &  \text{if $\psi\le (1+\sigma_u^2)|k|$}; \\
 \frac{(1+\sigma_u^2)k^*}{\psi} & \text{if $\psi> (1+\sigma_u^2)|k|$}.
  \end{cases}
\end{equation}
Indeed, when $\psi\le |k|(1+\sigma_u^2)$, letting $\gamma^2\to
\sigma_u^2+2-2|k|(1+\sigma_u^2)+\psi$ in~(\ref{eq:9}) produces
$\Theta=k/|k|$ as the unique admissible value of $\Theta$. As a result,
(\ref{eq:46}) reduces to $H_{11}=k^*/|k|$. 
Likewise, when $\psi>|k|(1+\sigma_u^2)$ and $\gamma^2\to
\sigma_u^2+2-\frac{|k|^2(1+\sigma_u^2)^2}{\psi}$,  
(\ref{eq:9}) holds for any $\Theta(s)$. Also, $\Upsilon_2\to
0$, and~(\ref{eq:67.LFT}) has the limit
$-\frac{e^{-i\varphi}\Upsilon_3}{\psi^{1/2}\Upsilon_1}$, which yields   
$H_{11}=\frac{(1+\sigma_u^2)k^*}{\psi}$. Remarkably, these are 
exactly the values of $H_{11}$ which we obtain by minimizing 
the expression for $P_e(i\omega,H_{11})$ in~(\ref{eq:121ps})
directly, proving that~(\ref{eq:23}) is, in fact, the
  identity. To demonstrate this, we present the following
    proposition which computes the value $\gamma_\circ'$ of the auxiliary
    Problem~\ref{P1a} and the corresponding minimizer exactly, using the
    method of Lagrange multipliers. This proposition shows that this value
    is precisely the same as the limit of the expression on the right hand side
    of~(\ref{eq:23}). This leads to the conclusion that
    $\gamma_\circ=\gamma_\circ'$ in this example; i.e., there is no gap
    between the value of the underlying optimal passive optimization
    problem~(\ref{eq:6'}) and the solution obtained using
    Proposition~\ref{Prop1.J} based on Theorem~\ref{T2}.          
A special case of this proposition appeared in~\cite{UJ2a}.  

\begin{proposition}\label{Prop1}
Consider the auxiliary optimal filtering problem~(\ref{eq:6''}) for the static
channel~(\ref{eq:1.bs}):  
\begin{eqnarray}
  (\gamma_\circ')^2=&&\inf P_e(i\omega,H_{11}) \nonumber \\
  && \mbox{subject to }
  \label{eq:15}
  |H_{11}(i\omega)|^2\le 1.
\end{eqnarray}

The following alternatives hold:
\begin{enumerate}[(a)]
\item
If $\psi\le (1+\sigma_u^2)|k|$, then the infimum in (\ref{eq:15}) is
achieved at $H_{11}$ given in the first line on the right hand side
of~(\ref{eq:5}). In this case, any passive equalizer
$\Xi(s)=\Delta(H(s),0)$, with $H(s)$ of the form 
\begin{equation}
  \label{eq:50}
  H(s)=
  \left[
    \begin{array}{cc}
      k^*/|k| & 0 \\ 0 & U_{22}(s)
    \end{array}
  \right], 
\end{equation}
where $U_{22}(s)$ is an arbitrary paraunitary transfer function, is an
optimal equalizer for the optimal passive equalization
problem~(\ref{eq:6'}).    

\item
On the other hand, when $\psi> (1+\sigma_u^2)|k|$, 
the infimum in (\ref{eq:15}) is achieved at $H_{11}$ given in the second
line on the right hand side of~(\ref{eq:5}).   
In this case, any passive equalizer
$\Xi(s)=\Delta(H(s),0)$, with $H(s)$ of the form 
\begin{equation}
  \label{eq:50.2}
\hspace{-.5cm}  H(s)=
  \left[
    \begin{array}{cc}
    \frac{(1+\sigma_u^2) k^*}{\psi}   & \frac{\sqrt{\psi^2-(1+\sigma_u^2)^2 |k|^2}}{\psi}U_{12}(s) \\ \frac{\sqrt{\psi^2-(1+\sigma_u^2)^2 |k|^2}}{\psi}U_{21}(s)~  & -\frac{(1+\sigma_u^2) k}{\psi}U_{12}(s)U_{21}(s)
    \end{array}
  \right], 
\end{equation}
where $U_{12}(s)$, $U_{21}(s)$ are arbitrary paraunitary transfer
functions, is a solution of the optimal passive equalization
problem~(\ref{eq:6'}). 
\end{enumerate}
The expression on the right-hand side of
equation~(\ref{eq:23}) gives the corresponding expressions for the optimal 
error power spectrum density:
\begin{equation}
\label{eq:7}
(\gamma_\circ')^2= \gamma_\circ^2=\begin{cases}
\psi-2(1+\sigma_u^2)|k|+(2+\sigma_u^2), \\
& \hspace{-3cm} \text{if $\psi\le (1+\sigma_u^2)|k|$}; \\
(2+\sigma_u^2)-\frac{(1+\sigma_u^2)^2|k|^2}{\psi}, \\ 
& \hspace{-3cm} \text{if $\psi> (1+\sigma_u^2)|k|$}.
  \end{cases}
\end{equation}
\end{proposition}

\emph{Proof: }
Since the coefficients of the objective function~(\ref{eq:121ps})
are constant, it will suffice to carry out
optimization in~(\ref{eq:15}) over the closed unit disk 
$\{H_{11},|H_{11}|\le 1\}$. The corresponding cost is independent of
$\omega$, and we will write it as $P_e(H_{11})$ in lieu of
$P_e(i\omega,H_{11})$.      

To prove claim (a), suppose first that $\psi<(1+\sigma_u^2)|k|$. 
Consider the Lagrangian function with the multiplier $\lambda\ge 0$
\begin{eqnarray*}
  \mathcal{L}&=& P_e(H_{11})-\lambda(1-|H_{11}|^2).
\end{eqnarray*}
The KKT optimality conditions are 
\begin{eqnarray}
  \label{eq:49}
 && (\psi+\lambda)H_{11}=(1+\sigma_u^2) k^*, \\ 
  \label{opt.con.complementarity}
&&\lambda(|H_{11}|^2-1)=0. 
\end{eqnarray} 
Based on the complementarity condition (\ref{opt.con.complementarity}), the
following two cases must be considered.

Case 1. $\lambda> 0$. In this case, the complementarity condition
(\ref{opt.con.complementarity}) yields $|H_{11}|=1$. Therefore,
(\ref{eq:49}) yields the critical value $\lambda=(1+\sigma_u^2)|k|-\psi$
which is positive under the assumption $\psi<(1+\sigma_u^2) |k|$. Then, the
corresponding minimizer is $H_{11}=k^*/|k|$. 

Case 2. $\lambda=0$. In this case, the optimality condition~(\ref{eq:49}) 
implies $H_{11}=\frac{(1+\sigma_u^2) k^*}{\psi}$.
However, this value of $H_{11}$ cannot be an optimal point
because $|H_{11}|>1$ when $\psi<(1+\sigma_u^2) |k|$. Thus, the 
solution to the problem~(\ref{eq:15}) is $H_{11}$ obtained in Case~1.

When $\psi= (1+\sigma_u^2)|k|$, the 
function $P_e$ reduces to
\begin{eqnarray*}
  P_e(H_{11})&=&(1+\sigma_u^2)\left|\sqrt{|k|}H_{11}-\frac{k^*}{\sqrt{|k|}}\right|^2
\nonumber \\
&& +(2+\sigma_u^2)-(1+\sigma_u^2) |k|; 
\end{eqnarray*}
it achieves minimum at $H_{11}=k^*/|k|$. 
 
Thus we observe that in both cases, the minimum in~(\ref{eq:15}) is achieved at
$H_{11}=k^*/|k|$. To obtain the corresponding coherent equalizer, we refer
directly to equations (\ref{eq:37}), since $H_{11}$ is physically
realizable on its own, and the transfer function $X_2$ in
condition (H3) of Theorem~\ref{two-step} is 0. As explained in
Remark~\ref{r=0}, in this situation additional noise channels are not
required to ensure the physical realizability of the filter. For
mathematical consistency, we can let 
$H_{12}=H_{21}=0$, and select $H_{22}(s)$ to be an arbitrary paraunitary
transfer function $U_{22}(s)$.   
This completes the proof of claim~(a). 

The proof of claim (b) proceeds in a similar manner. We again 
analyze the optimality
conditions~(\ref{eq:49}),~(\ref{opt.con.complementarity}). This time 
however, $\lambda=(1+\sigma_u^2)|k|-\psi$  
fails to be nonnegative since $\psi> (1+\sigma_u^2) |k|$. On the other hand, in
the case $\lambda=0$, we obtain that the minimum is achieved at
$H_{11}=\frac{(1+\sigma_u^2) k^*}{\psi}$ since
this value of $H_{11}$ satisfies the condition $|H_{11}|\le 1$. The
remaining entries of the optimal filter matrix $H(s)$ are obtained using
Theorem~\ref{two-step}. \hfill$\Box$

\begin{remark}\label{rem.bs.opt}
In order to obtain an optimal equalizer in this
example, it suffices to select a constant $\Theta(s)$. As we have shown,
when $\psi\le |k|(1+\sigma_u^2)$, the optimal equalizer is obtained using
$\Theta=k/|k|$, and when $\psi\le |k|(1+\sigma_u^2)$, a transfer
function $\Theta(s)$ can be selected arbitrarily. In the latter case,
choosing a dynamic parameter $\Theta(s)$ in~(\ref{eq:46}) delivers no
benefit, compared with choosing a constant parameter. 
\end{remark}

It is interesting to compare the optimal points of the constrained
optimization problem~(\ref{eq:15}) with optimal points of the
corresponding unconstrained 
optimization problem~(\ref{eq:71.nc}). When $\psi\le (1+\sigma_u^2) |k|$,
claim (a) of 
Proposition~\ref{Prop1} shows that the solutions to these two problems are
different. In the constrained problem~(\ref{eq:15}) the minimum is achieved  
on the boundary of the unit disk at $H_{11}=k^*/k$, whereas
the minimum of the unconstrained problem~(\ref{eq:71.nc}) is achieved
outside the unit disk, at 
$H_{11,*}=\frac{(1+\sigma_u^2) k^*}{\psi}$. The minimum value of the
problem~(\ref{eq:71.nc}) is $\gamma_*^2=2+\sigma_u^2-\frac{(1+\sigma_u^2)^2
  |k|^2}{\psi}<\gamma_\circ^2$.  
On the other hand, when $\psi> (1+\sigma_u^2)|k|$, claim (b) of
Proposition~\ref{Prop1} states that the two 
solutions are identical, and $\gamma_\circ^2=\gamma_*^2$. This situation
was envisaged in Section~\ref{S-proc}, and we now show that the threshold
condition $\psi> (1+\sigma_u^2)|k|$ which characterizes alternative (b)
can be obtained directly from Theorem~\ref{SDP.primal.LMI}. 

\begin{proposition}\label{SDP.primal.LMI.BS}
  $\psi> (1+\sigma_u^2) |k|$ if and only if there exists $\theta>0$ such that
  \begin{equation}
    \label{eq:16}
    \theta 
    \left[
      \begin{array}{cc}
      \psi  &-(1+\sigma_u^2) k \\
      -(1+\sigma_u^2) k^* & (2+\sigma_u^2)-\gamma_*^2  
      \end{array}
    \right]> 
    \left[
      \begin{array}{rr}
        1 & 0 \\
       0 & ~-1
      \end{array}
    \right].
  \end{equation}
\end{proposition}

\emph{Proof: }
Since $\gamma_*^2=2+\sigma_u^2-\frac{(1+\sigma_u^2)^2|k|^2}{\psi}$ and
$\theta>0$, (\ref{eq:16}) is equivalent to the inequality
\begin{equation}
  \label{eq:89}
    \left[
      \begin{array}{cc}
      \psi-\frac{1}{\theta}  &-(1+\sigma_u^2) k \\
      -(1+\sigma_u^2) k^* & \frac{1}{\theta}+\frac{(1+\sigma_u^2)^2|k|^2}{\psi}
      \end{array}
    \right]> 0. 
\end{equation}
It is easily checked that a $\theta>0$ for which~(\ref{eq:89}) holds exists
if and only if $\psi-\frac{(1+\sigma_u^2)^2|k|^2}{\psi}>0$. The latter condition
is equivalent to the inequality $\psi> (1+\sigma_u^2) |k|$.  
\hfill$\Box$

Proposition~\ref{SDP.primal.LMI.BS} and
  Theorem~\ref{SDP.primal.LMI} show that when $\psi> (1+\sigma_u^2) |k|$
  the constraint~(\ref{eq:15}) is inactive. This observation has
  an interesting interpretation, since the inequality
$\psi> (1+\sigma_u^2) |k|$ is equivalent to $\sigma_w^2> \bar
\sigma_w^2=\frac{(1+\sigma_u^2)|k|-\sigma_u^2|k|^2}{|m|^2}$. The latter
inequality sets a threshold on the intensity of the field $w$.
When the 
input field $w$ exceeds this threshold, the optimal filter is able to mix the
fields $y$ and $z$ in such a way that the intensity of the equalization error 
$e=\hat u-u$ is reduced, compared with the intensity of the error
$y-u$. The latter would be incurred if the equalizer was not used. Indeed,
the difference  
between the power spectrum densities of these two errors is
\begin{eqnarray*}
P_{y-u}-\gamma_\circ^2
      &=&\frac{|\psi-(1+\sigma_u^2) k|^2}{\psi} >0. 
\end{eqnarray*}
This shows that the equalizer is able to offset the high intensity field
$w$ by redirecting a fraction of this field to the output $\hat z$, and
`trade' it for the low intensity noise $z$. Note that the gap between
$P_{y-u}$ and the optimal power spectrum density $\gamma_\circ^2$ of the
equalization error increases as $\sigma_w^2$ increases.  

On the other hand, when $\sigma_w^2\le \bar \sigma_w^2$, the improvement is
marginal. It does not depend on $\sigma_w^2$:
\begin{eqnarray*}
P_{y-u}-\gamma_\circ^2
      &=&2(1+\sigma_u^2)(|k|-\mathrm{Re}k).
\end{eqnarray*}
According to~(\ref{eq:5}), the action of the optimal equalizer in this case
is limited to phase correction, $\hat u=\frac{k^*}{|k|}y$.
In the worst case scenario, when $k$ is real, the filter simply passes the
unaltered input $y$ through.
In this worst case, $e=y-u$ and $\gamma_\circ^2=P_{y-u}$; i.e., the
optimal equalizer is unable to improve the mean-square error.  

This analysis shows that the
capacity of an optimal coherent
equalizers to respond to noise in the transmission channel is restricted
when the signal to thermal noise ratio $\sigma_u^2/\sigma_w^2$ in the
quantum transmission channel exceeds
\[
\frac{\sigma_u^2}{\bar\sigma_w^2}=\frac{|m|^2\sigma_u^2}{(1+\sigma_u^2)|k|-\sigma_u^2|k|^2}=\frac{|m|^2}{|k|\sigma_u^{-2}+|k|-|k|^2}.
\]
The benefits of equalization become tangible only 
when the ratio $\sigma_u^2/\sigma_w^2$ is sufficiently small. This situation differs strikingly from the
situation encountered in the classical mean-square equalization theory. We
conjecture that this 
phenomenon holds in general when the channel environment noise is thermal,
and the equalizer is passive, and that the condition~(\ref{eq:21}) sets a
corresponding threshold on the signal to thermal noise ratio.  

We conclude the example by comparing numerical results obtained
from the SDP problem~(\ref{eq:71.LMI.l})--(\ref{eq:14.l}) with the results
obtained directly using the method of Lagrange multipliers; see Proposition~\ref{Prop1}.  
\begin{figure}[t]
  \centering
  \includegraphics[width=0.85\columnwidth]{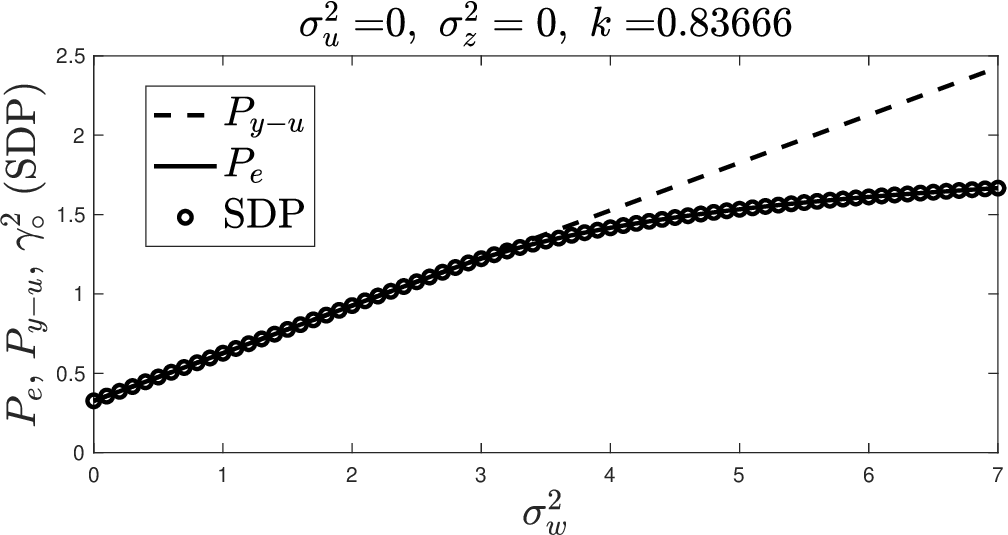}
  \caption{Power spectrum densities $P_{y-u}$ and $P_{e}$ (given
    by~(\ref{eq:7})) and the optimal
    value of the SDP problem
    (\ref{eq:71.LMI})--(\ref{eq:14}) for a range of $\sigma_w^2$, for the
  beamsplitter transmittance of $\eta=0.7$.}
  \label{fig.compare}
\end{figure}
\begin{figure}[t]
  \centering
  \includegraphics[width=0.85\columnwidth]{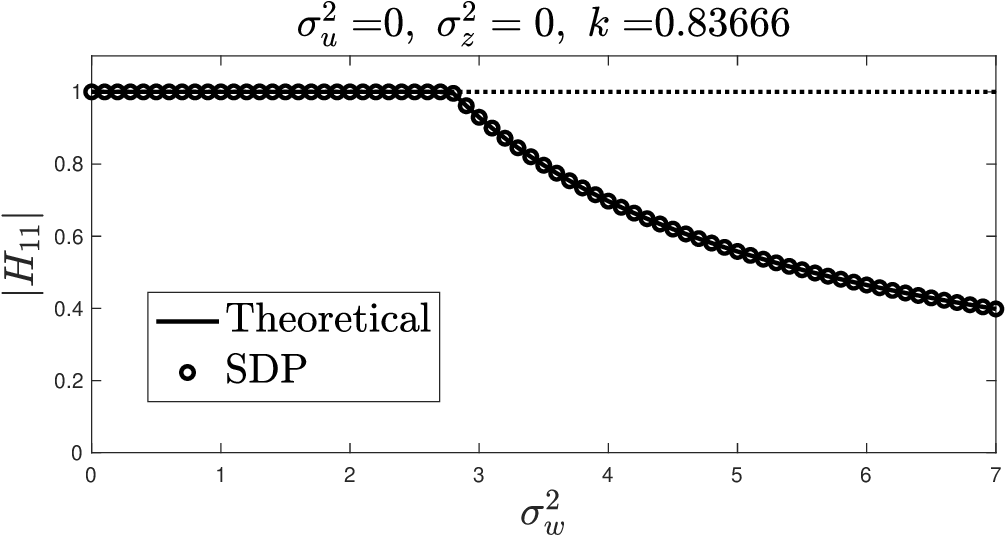}
  \caption{The theoretical (equation~(\ref{eq:5})) and numerically obtained
    (via the SDP problem (\ref{eq:71.LMI})--(\ref{eq:14})) optimal gains
    $|H_{11}|$ for a range of $\sigma_w^2$, for the beamsplitter transmittance of $\eta=0.7$.} 
  \label{fig.H11}
\end{figure}
For this comparison, consider a
quantum-mechanical beamsplitter as a special case of a static two-input
two-output quantum channel. In this case, $k=\sqrt{\eta}$,
$m=\sqrt{1-\eta}$, where $\eta\in(0,1)$ is the transmittance of the
device. That is, $y=\sqrt{\eta}u+\sqrt{1-\eta}w$.

Figure~\ref{fig.compare} shows the plot of the optimal value of
the LMI problem~(\ref{eq:71.LMI.l})--(\ref{eq:14.l})
obtained for this example numerically. Since the parameters of the
system are constant w.r.t. $\omega$, interpolation is not required in this
example, and one can use the obtained numerical value $H_{11}$ directly to
obtain an optimal equalizer, as was done in Proposition~\ref{Prop1}. 
The optimal $H_{11}$ obtained numerically is
real, the graph of the optimal $|H_{11}|$ for this example is shown in
Figure~\ref{fig.H11}. For comparison, Figures~\ref{fig.compare}
and~\ref{fig.H11} also show the graphs of the optimal $P_e$ in
equation~(\ref{eq:7}) and
the optimal gain $H_{11}$ obtained in
Proposition~\ref{Prop1}. Remarkably, both the graphs of the optimal value of
the optimization problem and the graphs of the optimal gain are essentially
identical to the corresponding graphs obtained directly using the
  method of Lagrange multipliers.

The threshold on the intensity of the noise $w$ separating the two
alternative equalization 
strategies can also be seen vividly in the graphs. 
With the chosen parameters, the threshold is
$\bar\sigma_w^2=\frac{(1+\sigma_u^2)\sqrt{\eta}-\sigma_u^2\eta}{1-\eta}$. When
$\sigma_w^2$ is below this threshold, the equalizing filter is given
by~(\ref{eq:50}), and we let $H_{22}(s)=1$:
\[
\hat u=y=\sqrt{\eta}u+\sqrt{1-\eta}w. 
\]
I.e., the optimal equalization policy is to pass the channel output $u$
unaltered. On the other hand, when $\sigma_w^2>\bar\sigma_w^2$, the optimal
equalizing filter is given by~(\ref{eq:50.2}). Letting $U_{12}=U_{21}=1$
yields the following expression for the mean-square optimal estimate of
$u$,     
\begin{eqnarray*}
\hat
  u&=&\frac{(1+\sigma_u^2)\eta}{\psi}u+\frac{(1+\sigma_u^2)\sqrt{\eta}\sqrt{1-\eta}}{\psi}w
       \\
&&+\frac{1}{\psi}\sqrt{\psi^2-(1+\sigma_u^2)^2\eta}z.
\end{eqnarray*} 
When $\sigma_w^2>\bar\sigma_w^2$,
$\frac{(1+\sigma_u^2)\sqrt{\eta}\sqrt{1-\eta}}{\psi}< \sqrt{1-\eta}$, i.e., 
the optimal filter applies a reduced gain to the input $w$, compared
with the gain $\sqrt{1-\eta}$ of the 
corresponding term in the expression for $y$. This results in the lower
intensity of the filtering error; see Figure~\ref{fig.compare}.
The figure confirms that when the intensity of the thermal noise $w$ is
sufficiently large, the optimal equalizer is
able to reduce the degrading effect of the auxiliary noise $w$ by trading
it for a smaller intensity noise injected through the channel $z$. On 
the contrary, when the noise $w$ has low intensity, such trade-off
is not possible, and the filter resorts to passing the channel output $y$
through without any modification. 

We conclude the example by pointing out that for a beamsplitter of
transmittance $\eta$, the optimal
equalizer~(\ref{eq:50.2}) can be implemented using a single beamsplitter with
the transmittance
$\frac{(1+\sigma_u^2)^2\eta}{(\sigma_u^2\eta+\sigma_w^2(1-\eta))^2}$. 
I.e., 
the optimal channel-equalizer system has the configuration shown in
Fig.~\ref{fig:bsplusbs}.

\subsection{Equalization of an optical cavity system}\label{ex2.revisited}

\subsubsection*{Guaranteed cost equalization} 

\begin{figure}[t]
  \begin{center}
\psfragfig[width=0.9\columnwidth]{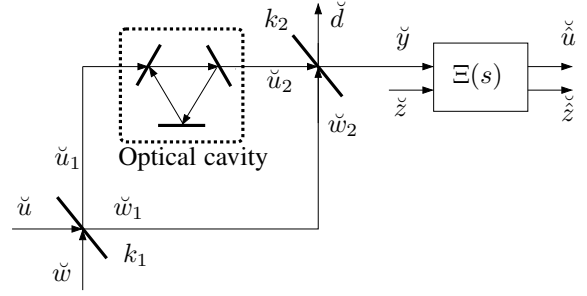}{ 
  \psfrag{Optical cavity}{Optical cavity}
  \psfrag{b}{\hspace{-1ex}$\breve w$}
  \psfrag{v}{$\breve u$}
  \psfrag{bp}{$\breve u_1$}
  \psfrag{vp}{$\breve w_1$}
  \psfrag{bp2}{$\breve u_2$}
  \psfrag{vp2}{$\breve w_2$}
  \psfrag{alpha}{$k_1$}
  \psfrag{beta}{$k_2$}
  \psfrag{w}{\hspace{-3ex}$w$}
  \psfrag{z}{$\breve z$}
  \psfrag{bhat}{$\breve{\hat u}$}
  \psfrag{wout}{$y_w$}
  \psfrag{vout}{$\breve{d}$}
  \psfrag{bout}{}
  \psfrag{zhat}{$\breve{\hat z}$}
  \psfrag{y}{$\breve{y}$}
  \psfrag{H}{\hspace{-2ex}$\Xi(s)$}}
  \caption{A cavity, beamsplitters and an equalizer system.}
  \label{cavity}
\end{center}
\end{figure}

  Consider the equalization system shown in Fig.~\ref{cavity}. The channel 
  consists of an optical cavity and two optical beamsplitters. As in
  the previous example, the input fields $u$ and $w$ are
  scalar. For simplicity, suppose that the transmittance parameters
  $k_1^2$, $k_2^2$ of the beamsplitters are equal, and that
    $k_1, k_2$ are real positive numbers, and so 
  $k_1=k_2=k$ is a real constant. Thus, the relations between the input and
    output fields of the beamsplitters are
    \begin{eqnarray*}
     && \left[
        \begin{array}{c}
   u_1 \\ w_1       
        \end{array}
      \right]=
      \left[
        \begin{array}{cc}
          k& m \\ -m & k
        \end{array}
      \right]     
      \left[
        \begin{array}{c}
   u \\ w       
        \end{array}
      \right], \quad
     \left[
        \begin{array}{c}
   y \\ d       
        \end{array}
      \right]=
      \left[
        \begin{array}{cc}
          k& m \\ -m & k
        \end{array}
      \right]     
      \left[
        \begin{array}{c}
   u_2 \\ w_2       
        \end{array}
      \right],  
    \end{eqnarray*}
$m=\sqrt{1-k^2}$.
    The  transfer function of the optical cavity is
$G_c(s)=\frac{s-\kappa+i\Omega}{s+\kappa+i\Omega}$, i.e., $u_2=G_c(s)u_1$. $\kappa>0$, $\Omega$
are real numbers; $2\kappa$ is the coupling constant which
  characterizes coupling between the input field $u_1$ and the cavity mode, and
  $\Omega$ is the detuning of the cavity
  frequency from the reference frequency; c.f.~(\ref{eq:new-cavity-simple})
where $\Omega=0$ was assumed.         
Then the elements 
  of the transfer function $G(s)$ of the channel are  
  \begin{eqnarray}
    \label{eq:29}
G_{11}(s)&=&k^2G_c(s)-(1-k^2), \nonumber \\
G_{12}(s)&=&k\sqrt{1-k^2}(G_c(s)+1), \nonumber \\
G_{21}(s)&=&-k\sqrt{1-k^2}(G_c(s)+1), \nonumber \\
G_{22}(s)&=&k^2-(1-k^2)G_c(s).
  \end{eqnarray}
Our standing assumptions in this section are that $\sigma_w^2>\sigma_u^2>0$
and $k^2<\frac{1}{2}$. 
Under these assumptions,
\begin{eqnarray}
&& \rho\triangleq
   1+\frac{\sigma_u^2}{2(\sigma_w^2-\sigma_u^2)k^2(1-k^2)}>1, \quad
   \hat\rho\triangleq \frac{\rho-1}{\rho+1}\in(0,1), \nonumber \\
&&  \delta\triangleq \frac{\sqrt{1-k^2}}{k}>1, \quad \hat\delta\triangleq
   \frac{\delta^2+1}{\delta^2-1}=\frac{1}{1-2k^2}>1.  
\label{eq:10} 
\end{eqnarray}

From~(\ref{eq:90}), we have 
\begin{equation}
  \label{eq:96}
\Upsilon_3=\sqrt{\sigma_u^2+2-\gamma^2}>0.
\end{equation}
In the next proposition, the following notation will be used:
\begin{eqnarray}
 \beta&\triangleq&\frac{1+\sigma_u^2}{\Upsilon_3\sqrt{2(\sigma_w^2-\sigma_u^2)(1+\rho)}}(\delta-\frac{1}{\delta}), \nonumber \\
  \alpha&\triangleq&\sqrt{\beta^2-1},
  \quad
  \nu\triangleq\sqrt{\beta^2\hat\delta^2-\hat\rho}, \label{eq:11} \\
\mu&\triangleq& \sqrt{2(\sigma_w^2-\sigma_u^2)k^2(1-k^2)(1+\rho)},
                \nonumber \\
N_1(s)&=&\beta(s+i\Omega)+\beta\hat\delta\kappa,
          \nonumber \\
N_2(s)&=&\alpha(s+i\Omega)+\nu\kappa. \nonumber 
\end{eqnarray}

\begin{proposition}\label{Prop.cav}
Suppose $\gamma$ is chosen so that $\beta>1$ and $\gamma^2<\sigma_u^2+2$. Then
$H_{11}(s)\in\mathcal{H}_{11,\gamma}$ if and only if 
\begin{eqnarray}
  \label{eq:46.1}
  H_{11}(s)&=&-\frac{\Upsilon_3}{\mu} 
\frac{s+\kappa+i\Omega}{N_1(s)-\Theta(s)N_2(s)},
\end{eqnarray}
where $\Theta(s)$ is a stable rational transfer function analytic in the
closed right half-plane, which satisfies $\|\Theta\|_\infty<1$ and the
frequency domain condition
\begin{eqnarray}
  \label{eq:9.1}
\left|\frac{N_1(i\omega)-\Theta(i\omega)N_2(i\omega)}{i(\omega+\Omega)+\kappa}
\right|\ge  \frac{\Upsilon_3}{\mu}  \quad \forall \omega\in\bar{\mathbf{R}}. 
\end{eqnarray}
\end{proposition}

\emph{Proof: }
Using the notation in~(\ref{eq:10}), the function $\Psi(s)$ given in
equation~(\ref{eq:47}) is expressed as
\begin{eqnarray*}
  \Psi(s)&=&\sigma^2_u+2(\sigma_w^2-\sigma_u^2)k^2(1-k^2)
   \left(1+\frac{(s+i\Omega)^2+\kappa^2}{(s+i\Omega)^2-\kappa^2}\right) \\
  \\
&=&2(\sigma_w^2-\sigma_u^2)k^2(1-k^2)(1+\rho)\frac{(s+i\Omega)^2-\hat\rho\kappa^2}{(s+i\Omega)^2-\kappa^2}.
\end{eqnarray*}
Clearly, $\Psi(s)$ has full normal rank. It admits the spectral
decomposition~(\ref{eq:61}) with the spectral factor\footnote{As in
  Proposition~\ref{Prop1.J}, one can 
use a spectral factor $M_1(s)=e^{i\varphi}M(s)$ in lieu of $M(s)$ in~(\ref{eq:54}). The
definitions of $\Upsilon_1(s)$, $\Upsilon_2(s)$ 
will then need to be updated accordingly, as was done in
Proposition~\ref{Prop1.J}.}       
\begin{equation}
  \label{eq:54}
  M(s)=\mu 
\frac{s+\kappa\sqrt{\hat\rho}+i\Omega}{s+\kappa+i\Omega}.
\end{equation}
Both $M(s)$ and $M^{-1}(s)$ are stable and
analytic in the half-plane
$\mathrm{Re}s>-\kappa\sqrt{\hat\rho}$. Using~(\ref{eq:54}), the transfer
function $Q(s)$ in~(\ref{eq:75}) is expressed as
\begin{eqnarray*}
  Q(s)=\frac{1+\sigma_u^2}{\sqrt{2(\sigma_w^2-\sigma_u^2)(1+\rho)}}(\delta-\frac{1}{\delta})\frac{s+\hat\delta\kappa+i\Omega}{s+\kappa\sqrt{\hat\rho}+i\Omega}.
\end{eqnarray*}
Using this information, one can readily check that the matrix $\Upsilon(s)$
in which 
\begin{eqnarray}
  \Upsilon_1(s)&=&\frac{N_1(s)}{s+\sqrt{\hat\rho}\kappa+i\Omega}
=\beta\frac{s+\hat\delta\kappa+i\Omega}{s+\sqrt{\hat\rho}\kappa+i\Omega},
                   \nonumber \\ 
\Upsilon_2(s)&=&\frac{N_2(s)}{s+\sqrt{\hat\rho}\kappa+i\Omega}
=\alpha\frac{s+\frac{\nu}{\alpha}\kappa+i\Omega}{s+\sqrt{\hat\rho}\kappa+i\Omega}, 
   \label{eq:95}
\end{eqnarray}
$\Upsilon_3$ is defined in~(\ref{eq:96}) and $\Upsilon_4=0$, is a
$J$-spectral factor of the corresponding matrix $\Phi(s)$. Indeed, when
$\beta>1$, the constants $\alpha$ and $\nu$ in~(\ref{eq:11}) are well
defined, and the identity~(\ref{eq:68}) can be verified directly.
Also, since $\sqrt{\hat\rho}\kappa>0$,
$\Upsilon_1(s)$ and $\Upsilon_2(s)$ are stable and are analytic in the
half-plane $\mathrm{Re}(s)>-\sqrt{\hat\rho}\kappa$. Therefore, 
$\Upsilon(s)$ also has these properties. Similarly, since  
$\hat\delta$ is positive, then $\Upsilon_1(s)^{-1}$ is also stable and is
analytic in the half-plane $\mathrm{Re}(s)>-\hat\delta\kappa$. 

Finally, we note that
\begin{equation}
  \label{eq:97}
\Upsilon(s)^{-1}= \frac{1}{K(s)}
\left[
  \begin{array}{cc}
    0 & -\Upsilon_2(s) \\
-\Upsilon_3 & \Upsilon_1(s)
  \end{array}
\right], 
\end{equation}
where 
\begin{equation}
  \label{eq:99}
  K(s)=\det\Upsilon(s)
=-\Upsilon_3\Upsilon_2(s)=-\Upsilon_3 \alpha
\frac{s+\frac{\nu}{\alpha}\kappa+i\Omega}{s+\sqrt{\hat\rho}\kappa+i\Omega}. 
\quad 
\end{equation}
Therefore $\frac{1}{K(s)}$ is stable and is analytic in the half-plane
$\mathrm{Re}s> -\frac{\nu}{\alpha}\kappa$. Thus we conclude that
$\Upsilon(s)^{-1}$ is stable and is analytic in a half-plane
$\mathrm{Re}s>-\tau$, $\exists\tau>0$.

These properties verify the conditions of Theorem~\ref{T2}. Therefore
$H_{11}(s)\in\mathcal{H}_{11,\gamma}$ if and only if there
  exists a transfer function $\Theta(s)$ with properties described in that
  theorem for which $H_{11}(s)$ can be expressed by
equation~(\ref{eq:67}).

Since we chose $\Upsilon_4=0$, we can use
Corollary~\ref{cor.LFT} to obtain the general
form of a feasible $H_{11}(s)$. Substituting~(\ref{eq:54}),~(\ref{eq:95}) 
in~(\ref{eq:67.LFT}) yields~(\ref{eq:46.1}). The
frequency domain condition~(\ref{eq:9.1}) follows from~(\ref{eq:79})
and~(\ref{eq:45.LFT}) in the same manner.    
\hfill$\Box$   

As in the previous section, it is useful to derive sufficient
conditions which would allow us to obtain a $\Theta(s)$ which
solves~(\ref{eq:9.1}). For this, we
restrict attention to constant $\Theta$'s. 

\begin{corollary}\label{cav.suffcond}
Under the conditions of Proposition~\ref{Prop.cav}, if 
\begin{equation}
  \label{eq:9.4}
\beta+\alpha>\mu^{-1}\sqrt{\sigma_u^2+2-\gamma^2},
\end{equation} 
then for any constant $\Theta\in
(-1,\min\{\alpha^{-1}(\beta-\mu^{-1}\Upsilon_3),0\})$ the corresponding
$H_{11}(s)$ given by equation~(\ref{eq:46.1}) belongs to 
$\mathcal{H}_{11,\gamma}$.  
\end{corollary}

\emph{Proof: }
For a constant $\Theta$, (\ref{eq:9.1}) is equivalent to
\begin{equation*}
  \label{eq:9.2}
\frac{\Upsilon_3}{\mu} 
\sqrt{\frac{(\omega+\Omega)^2+\kappa^2}{(\beta-\Theta\alpha)^2(\omega+\Omega)^2
+(\beta\hat\delta-\Theta\nu)^2}}
\le 1   \quad \forall \omega\in\bar{\mathbf{R}}. 
\end{equation*}
The maximum of the expression on the left-hand side is equal to
$\frac{\Upsilon_3}{\mu}/\min\{|\beta-\Theta\alpha|,|\beta\hat\delta-\Theta\nu|\}$. 
Therefore, (\ref{eq:9.1}) is equivalent to the condition 
\begin{equation}
  \label{eq:9.3}
\min\{|\beta-\Theta\alpha|,|\beta\hat\delta-\Theta\nu|\}\ge \mu^{-1} \sqrt{\sigma_u^2+2-\gamma^2}.
\end{equation}

Next, we show that~(\ref{eq:9.3}) holds for any $\Theta\in
(-1,\min\{\alpha^{-1}(\beta-\mu^{-1}\Upsilon_3),0\})$. Indeed, condition
(\ref{eq:9.4}) guarantees that this interval is not an empty set. Then  
for any $\Theta$ in that interval,
\[
\min\{|\beta-\Theta\alpha|,|\beta\hat\delta-\Theta\nu|\}=\beta-\Theta\alpha.
\]  
This identity holds because $\Theta<0$, $\beta\hat\delta>\beta$ and
$\nu>\alpha$ due to $\hat\delta>1$, $\hat\rho<1$. Furthermore,
$\Theta<\alpha^{-1}(\beta-\mu^{-1}\Upsilon_3)$ implies 
$
\beta-\Theta\alpha>\mu^{-1}\Upsilon_3.
$
This validates~(\ref{eq:9.3}) and~(\ref{eq:9.1}). The claim
then follows from Proposition~\ref{Prop.cav}.
\hfill$\Box$

Finally, we apply Corollary~\ref{cav.suffcond} and Theorem~\ref{two-step}
to obtain a complete physically 
realizable suboptimal equalizer for the cavity system in this example. For
convenience, we introduce the additional notation
\begin{eqnarray}
  \label{eq:22}
  a=-\frac{\Upsilon_3}{\mu(\beta-\Theta\alpha)}, \quad
  c=\frac{\beta\hat\delta-\Theta\nu}{\beta-\Theta\alpha}.
\end{eqnarray}
Conditions of
Corollary~\ref{cav.suffcond} allow us to select
$\Theta\in (-1,\min\{\alpha^{-1}(\beta-\mu^{-1}\Upsilon_3),0\})$, i.e.,
$\Theta<0$ and
\begin{eqnarray*}
  \label{eq:24}
  |a|&=&\frac{\Upsilon_3}{\mu|\beta-\Theta\alpha|}=\frac{\Upsilon_3}{\mu(\beta-\Theta\alpha)}<1,
  \\
  c&=&\frac{\beta\hat\delta-\Theta\nu}{\beta-\Theta\alpha}>1>|a|.
\end{eqnarray*}
Hence $c^2-a^2>0$, $1-a^2>0$.

Using this notation, the transfer function $H_{11}(s)$ in~(\ref{eq:46.1})
can be written in a compact form
\begin{eqnarray}
  \label{eq:18}
  H_{11}(s)&=&a\frac{s+\kappa+i\Omega}{s+c\kappa+i\Omega}.
\end{eqnarray}
Clearly, it satisfies condition (H1) of
Theorem~\ref{two-step}. The frequency domain
 condition~(\ref{eq:9.1}) ensures that condition (H2) is also satisfied. 
Then we compute
\begin{eqnarray}
  \label{eq:28}
  X_1(s)&=&X_2(s)=1-H_{11}(s)H_{11}(s)^H
\nonumber \\
&=&(1-a^2)\frac{(s+i\Omega)^2-\frac{c^2-a^2}{1-a^2}\kappa^2}{(s+i\Omega)^2-c^2\kappa^2}. 
\end{eqnarray}
It is easy to check that $X_1(s)$ and $X_2(s)$ are para-Hermitian and satisfy
condition (H3) of Theorem~\ref{two-step}. Let us define the spectral
factors of $X_1(s)$, $X_2(s)$, 
\begin{eqnarray}
  \label{eq:30}
   H_{12}(s)&=&-\sqrt{1-a^2}\frac{s+\sqrt{\frac{c^2-a^2}{1-a^2}}\kappa +i\Omega}{s+c\kappa+i\Omega},
  \\
\tilde H_{21}(s)&=&-H_{12}(s)=\sqrt{1-a^2}\frac{s+\sqrt{\frac{c^2-a^2}{1-a^2}}\kappa +i\Omega}{s+c\kappa+i\Omega},
  \nonumber \\
\tilde
  H_{21}^{-1}(s)&=&\frac{1}{\sqrt{1-a^2}}\frac{s+c\kappa+i\Omega}{s+\sqrt{\frac{c^2-a^2}{1-a^2}}\kappa +i\Omega},
\nonumber 
\end{eqnarray}
and also select 
\begin{equation*}
  \label{eq:63}
  U(s)=\frac{s-\sqrt{\frac{c^2-a^2}{1-a^2}}\kappa +i\Omega}{s+\sqrt{\frac{c^2-a^2}{1-a^2}}\kappa +i\Omega}. 
\end{equation*}
This transfer function is paraunitary, stable and analytical in the right
half-plane $\mathrm{Re}s>-\sqrt{\frac{c^2-a^2}{1-a^2}}$, as required by
Theorem~\ref{two-step}. 
Using these definitions the remaining blocks of $H(s)$ are obtained
according to~(\ref{eq:81}):
\begin{eqnarray}
  \label{eq:72}
  H_{21}(s)&=&U(s)\tilde H_{21}(s) \nonumber \\
&=&\sqrt{1-a^2}\frac{s-\sqrt{\frac{c^2-a^2}{1-a^2}}\kappa+i\Omega}{s+c\kappa+i\Omega},
    \nonumber \\
  H_{22}(s)&=&-U(s)(\tilde H_{21}^{-1}(s))^HH_{11}(s)^HH_{12}(s) \nonumber \\
&=&a\frac{s-\kappa+i\Omega}{s+c\kappa+i\Omega}.
\end{eqnarray}
The following proposition which follows from
  Theorem~\ref{two-step} summarizes our analysis. 
 
\begin{proposition}\label{Prop.cav.complete}
Given a constant $\gamma$ which satisfies the conditions of
Proposition~\ref{Prop.cav} and Corollary~\ref{cav.suffcond},
consider the transfer function $\Xi(s)=\Delta(H(s),0)$
where $H(s)$ is composed of the blocks defined in
equations~(\ref{eq:18}),~(\ref{eq:30}) and~(\ref{eq:72}); see~(\ref{eq:98a}).
Then $\Xi(s)=\Delta(H(s),0)$ is a passive
physically realizable stable and causal guaranteed cost equalizer for the
cavity system under consideration, and
\begin{equation}
  \label{eq:78}
  \sup_\omega P_e(i\omega,\Xi)<\gamma^2.
\end{equation}
\end{proposition}

It is worth pointing out that the guaranteed cost equalizer in this example
can be realized using an interconnection of an optical cavity and two
beamsplitters shown in Figure~\ref{fig:cavity.filter}.
The transfer function of the optical cavity in the figure is
\begin{equation*}
y_2=H_c(s)y_1, \quad  H_c(s)=\frac{s-c\kappa+i\Omega}{s+c\kappa+i\Omega},
\end{equation*}
and the beamsplitters' operators are
    \begin{eqnarray*}
     && \left[
        \begin{array}{c}
   y_1 \\ z_1       
        \end{array}
      \right]=
      \left[
        \begin{array}{cc}
        \xi_1 &  \eta_1 \\ \eta_1 & -\xi_1
        \end{array}
      \right]     
      \left[
        \begin{array}{c}
   y \\ z       
        \end{array}
      \right], \quad
     \left[
        \begin{array}{c}
   \hat u \\ \hat z       
        \end{array}
      \right]=
      \left[
        \begin{array}{cc}
          \eta_2& \xi_2 \\ \xi_2 & -\eta_2
        \end{array}
      \right]     
      \left[
        \begin{array}{c}
   y_2 \\ z_2       
        \end{array}
      \right],  
    \end{eqnarray*}
where
\begin{eqnarray*}
 && \eta_1=-\sqrt{\frac{c+a^2-\sqrt{(c^2-a^2)(1-a^2)}}{2c}}, \\
 && \xi_1=\sqrt{1-\eta_1^2}=\sqrt{\frac{c-a^2+\sqrt{(c^2-a^2)(1-a^2)}}{2c}}, \\
 && \eta_2=-\sqrt{\frac{c-a^2-\sqrt{(c^2-a^2)(1-a^2)}}{2c}}, \\
 && \xi_2=\sqrt{1-\eta_2^2}=\sqrt{\frac{c+a^2+\sqrt{(c^2-a^2)(1-a^2)}}{2c}}.
\end{eqnarray*}
\begin{figure}[t]
  \begin{center}
\psfragfig[width=0.7\columnwidth]{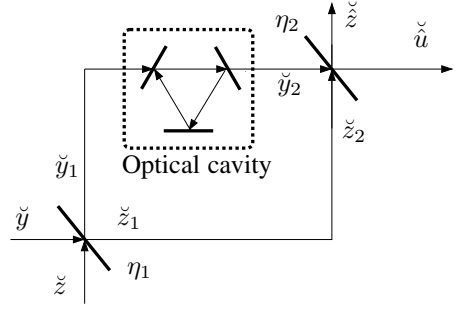}{ 
  \psfrag{Optical cavity}{Optical cavity}
  \psfrag{b}{\hspace{-1ex}$\breve z$}
  \psfrag{v}{$\breve y$}
  \psfrag{bp}{$\breve y_1$}
  \psfrag{vp}{$\breve z_1$}
  \psfrag{bp2}{$\breve y_2$}
  \psfrag{vp2}{$\breve z_2$}
  \psfrag{alpha}{$\eta_1$}
  \psfrag{beta}{$\eta_2$}
  \psfrag{w}{\hspace{-3ex}$w$}
  \psfrag{bhat}{$\breve{\hat u}$}
  \psfrag{wout}{$y_w$}
  \psfrag{vout}{$\breve{\hat z}$}
  \psfrag{y}{$\breve{\hat u}$}
  \psfrag{H}{\hspace{-2ex}$\Xi(s)$}}
  \caption{A cavity and beamsplitters realization of the equalizer.}
  \label{fig:cavity.filter}
\end{center}
\end{figure}

Proposition~\ref{Prop.cav.complete} reduces the question of finding a
suboptimal physically realizable equalizer to checking
whether~(\ref{eq:9.4}) is satisfied for a given 
$\gamma$ such that $\gamma<\sigma_u^2+2$ and $\beta>1$. It is also possible
to minimize the upper bound $\gamma^2$ on $\sup_\omega
P_e(i\omega,\Xi)$ over the set $\{\gamma\colon
  \gamma^2\in(0,\sigma_u^2+2),  
\beta>1,\beta+\alpha>\mu^{-1}\Upsilon_3\}$. This will lead to a
    suboptimal solution to the problem. Figure~\ref{fig:cav.subopt}
illustrates this. The solid line in
  Figure~\ref{fig:cav.subopt} shows such a suboptimal $\gamma^2$ obtained for a range of values of
$\sigma_w^2>\sigma_u^2$, where $\sigma_u^2=0.1$, $k=0.4$, $\kappa=0.5\times
10^9$,
$\Omega=10^9$.   
\begin{figure}[t]
  \centering
  \includegraphics[width=0.85\columnwidth]{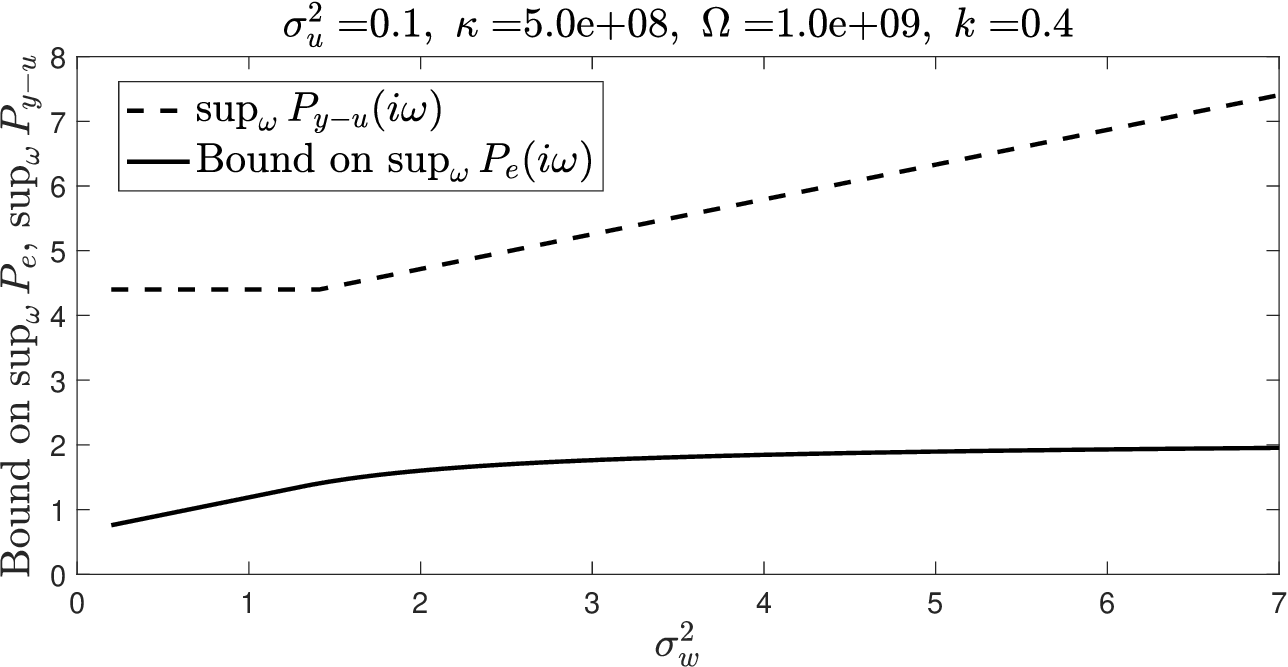}
  \caption{$\sup_\omega P_{y-u}(i\omega)$ (the dash line) and the optimized bound on
    $\sup_\omega P_{e}(i\omega)$ (the solid line)  
    for a range of $\sigma_w^2$, for an optical cavity.} 
  \label{fig:cav.subopt}
\end{figure}
For comparison, the figure shows the graph of the error power spectrum
density $\sup_\omega P_{y-u}$ of the system without an equalizer (the
dashed line). The
advantage of equalization is quite clear from this
figure. Fig.~\ref{fig:cav.Bode.subopt} shows the Bode plot of one of the
suboptimal transfer functions $H_{11}(s)=-\frac{s+5\times 10^8+10^9i}{s+7.961\times 10^8+10^9i}$
obtained for the cavity system with $\sigma_w^2=0.2$. It confirms that
$|H_{11}(i\omega)|\le 1$.  
\begin{figure}[t]
  \centering
  \includegraphics[width=0.85\columnwidth]{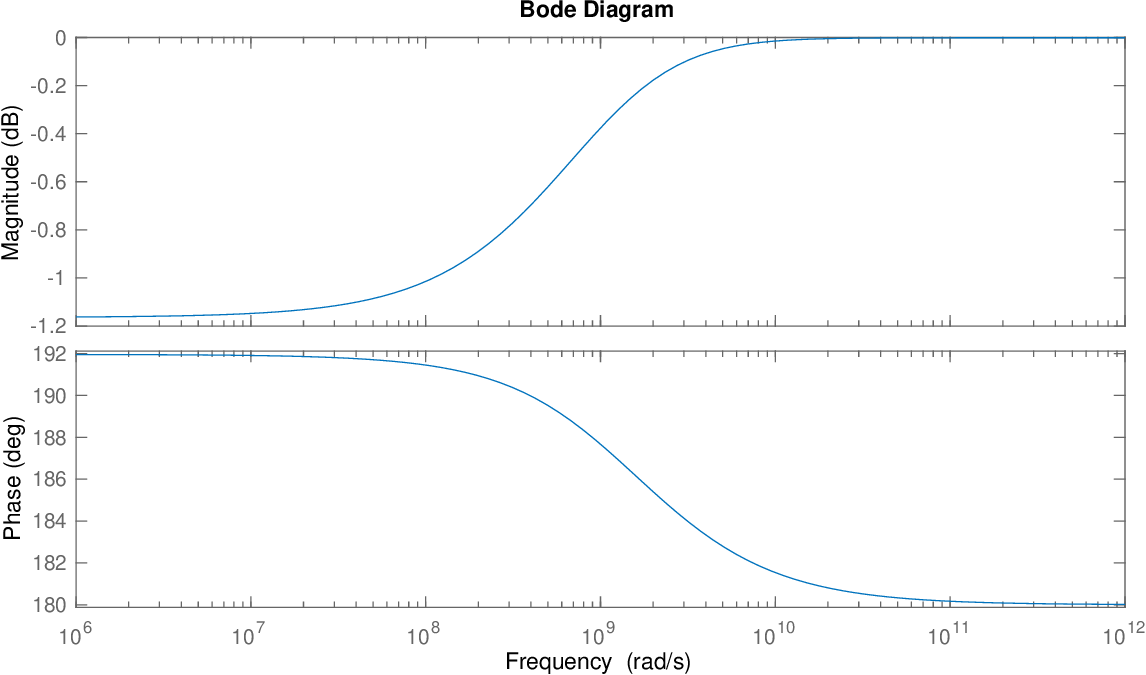}
  \caption{The Bode plot of one of the
suboptimal transfer functions $H_{11}(s)$
    obtained using equation~(\ref{eq:46.1}) with $\Theta=-0.9998$. The
    parameters of the system are $\sigma_w^2=0.2$, $\sigma_u^2=0.1$,
    $k=0.4$, $\kappa=0.5\times 10^9$, $\Omega=10^9$.} 
  \label{fig:cav.Bode.subopt}
\end{figure}

\subsubsection*{Equalization via semidefinite programming} 

We now illustrate the application of the approximation
technique presented in
Section~\ref{semidef}. For this, we consider the same cavity
system with  parameters
$k=0.4$, 
$\kappa=0.5\times 10^9$, $\Omega=10^9$, $\sigma_w^2=0.2$, $\sigma_u^2=0.1$. Recall that
the transfer function matrix $G(s)$ of that system is a $2\times 2$ matrix,
its elements are given in~(\ref{eq:29}).

To apply the algorithm described in Section~\ref{semidef} to this system,
first a set of $L=21$ 
points $\omega_l$ was selected which included $0$, 
ten logarithmically spaced frequency points in the interval
$[10^5,10^9]$ and the corresponding negative frequencies. With these
data, the LMI problem~(\ref{eq:71.LMI.l})--(\ref{eq:14.l}) was  
solved numerically and the array of values $H_{11,l}$ was obtained, $l=1,
\ldots, L$, along with the value of the optimization
  problem. In this example, this value was obtained to be 
$\tilde\gamma^2=0.7057$. It is worth noting
  that $\tilde\gamma^2<\sigma_u^2+2$; this validates~(\ref{eq:53}).
This set of data was then used to solve the Nevanlinna-Pick interpolation
problem. 

The procedure 
outlined in Section~\ref{semidef}  
involves mapping the half-plane $\mathrm{Re}s>-\tau$ conformally onto the
half-plane $\mathrm{Re}s>0$, performing interpolation over this half-plane
to obtain $\hat H_{11}(s)$, then obtaining $H_{11}(s)$ via~(\ref{eq:52}). To
implement this procedure, we used the conformal mapping
$s'=s+\tau$, where we let 
$\tau=10^5$ (i.e., a value of four orders of magnitude less than the
amplitude of the system pole). This ensured that the grid points
$i\omega_l$ on the  
imaginary axis were mapped conformally into the interior of the right
half-plane $\mathrm{Re}s>0$, as required by the algorithm in
Section~\ref{semidef}.

Theorem~NP in~\cite{Kovalishina-1984} allows to obtain a
solution using~(\ref{eq:58}), provided the Pick matrix $\mathbf{P}$ is  
positive definite. This assumption of the Nevanlinna-Pick interpolation
theory was satisfied in this example.
 
The method gives the analytical expression~(\ref{eq:58})
  for $\hat H_{11}(s)$. Since it involves inverting the Pick matrix $\mathbf{P}$, a
  closed form expression for~(\ref{eq:58}) is quite cumbersome, even when a
  modest number of grid points is selected. Therefore we validated our
  approach numerically. For this, we selected additional
grid points on the imaginary axis, while keeping the original frequencies as a
control set. Then we computed interpolated values of $H_{11}(i\omega)$ at
those grid points, using~(\ref{eq:58}),~(\ref{eq:52}) with
  $\Theta(s)=-0.95,0$ and $0.95$. Also, the 
corresponding normalized values of the error power spectrum density, 
$P_e(i\omega,H_{11})/{\tilde\gamma^2}$, were computed. The graphs of these
quantities are shown in Fig.~\ref{fig:psds} using solid lines.
%
%
\begin{figure}[t]
  \centering
  \begin{subfigure}{\columnwidth}
  \includegraphics[width=0.9\columnwidth]{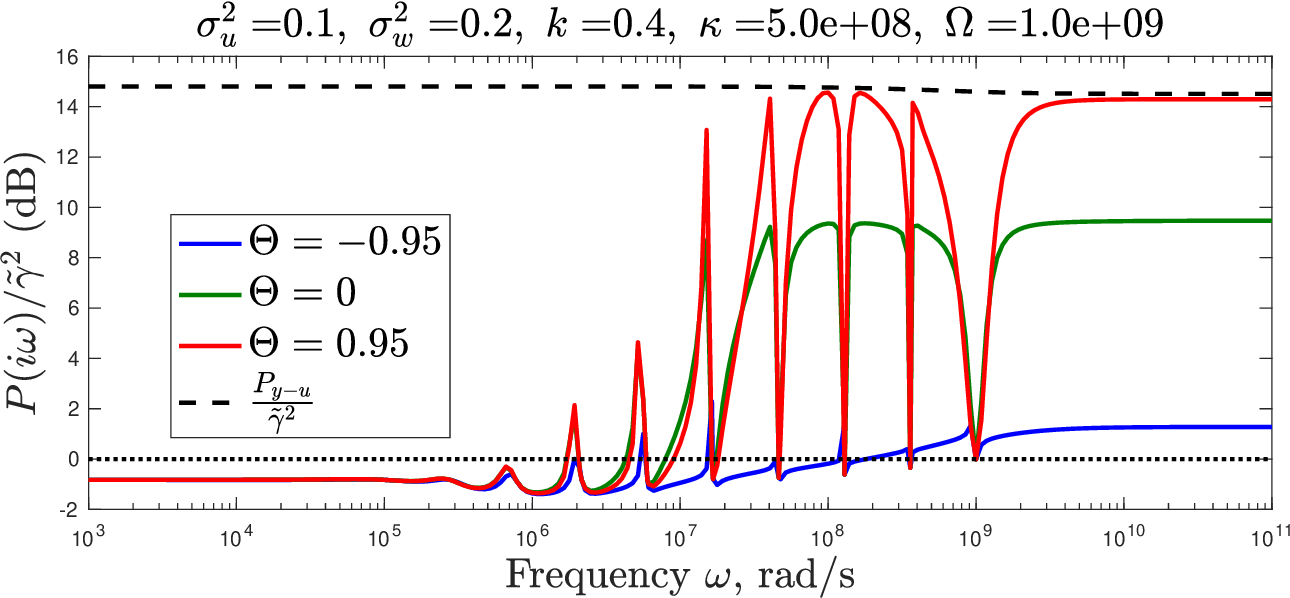}
  \caption{}\label{fig:psds}
\end{subfigure}
  \begin{subfigure}{\columnwidth}
  \includegraphics[width=0.9\columnwidth]{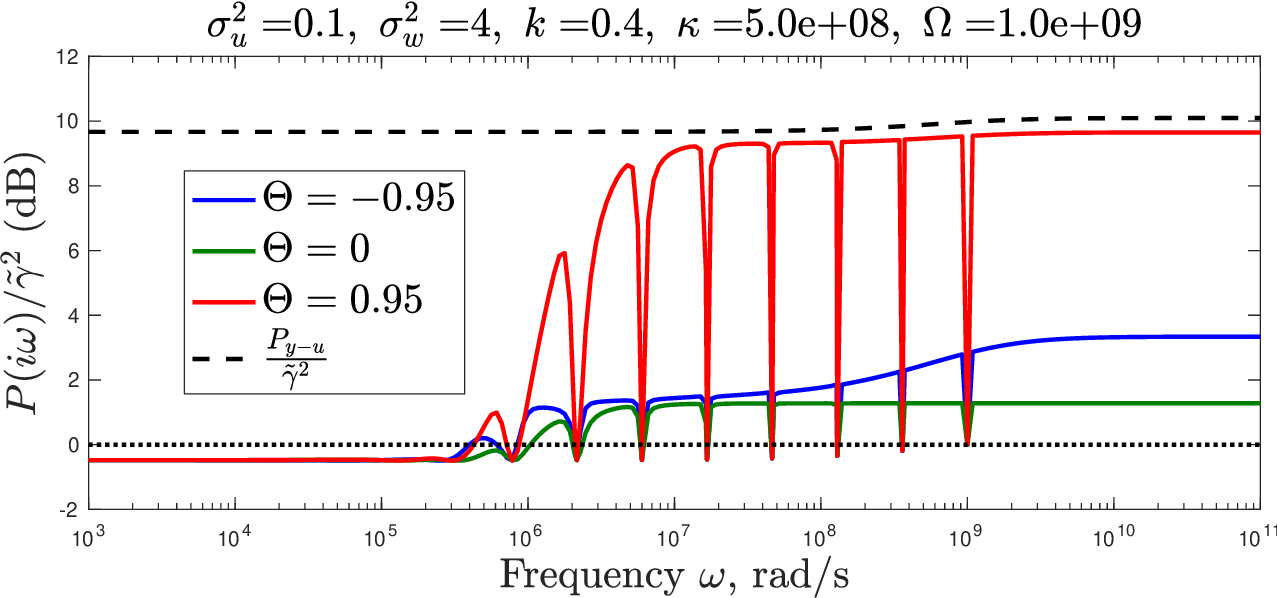}
  \caption{}\label{fig:psds.sig2w4}
\end{subfigure}
\caption{The normalized computed error power spectrum densities
    $P_e/{\tilde\gamma^2}$ (the solid lines) and the
    normalized error $P_{y-u}/{\tilde\gamma^2}$ (the dash line): (a)
    $\sigma_w^2=0.2$; (b) $\sigma_w^2=4$.}
  \label{fig:psds.all}
\end{figure}
\begin{figure}[t]
  \centering
  \begin{subfigure}{\columnwidth}
  \includegraphics[width=0.9\columnwidth]{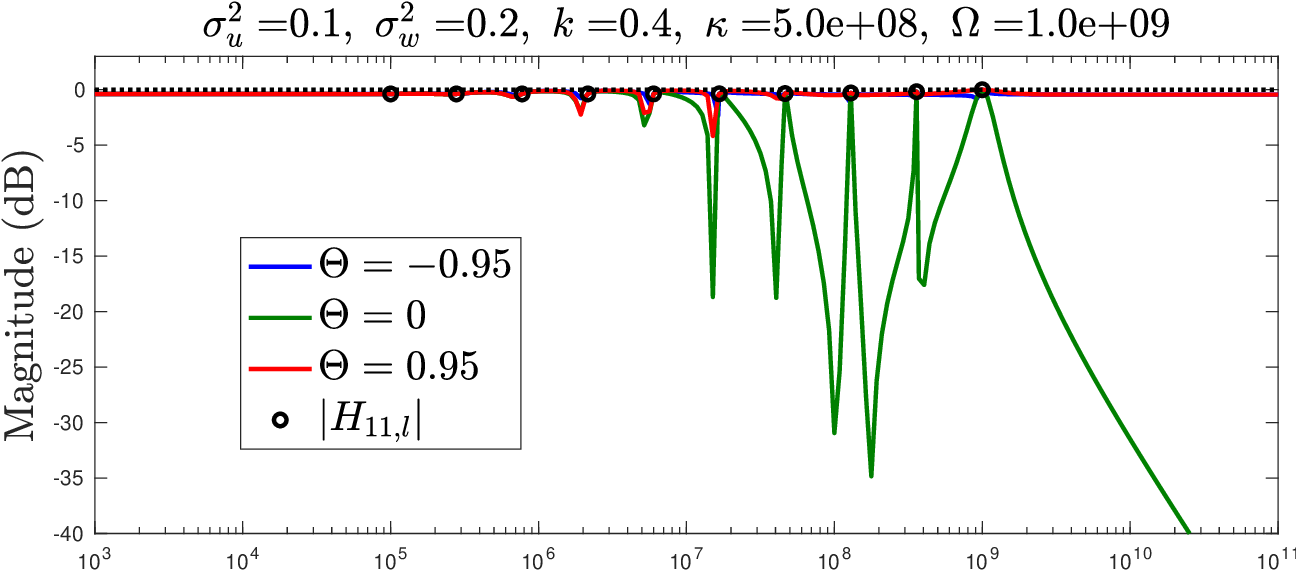}
  \includegraphics[width=0.9\columnwidth]{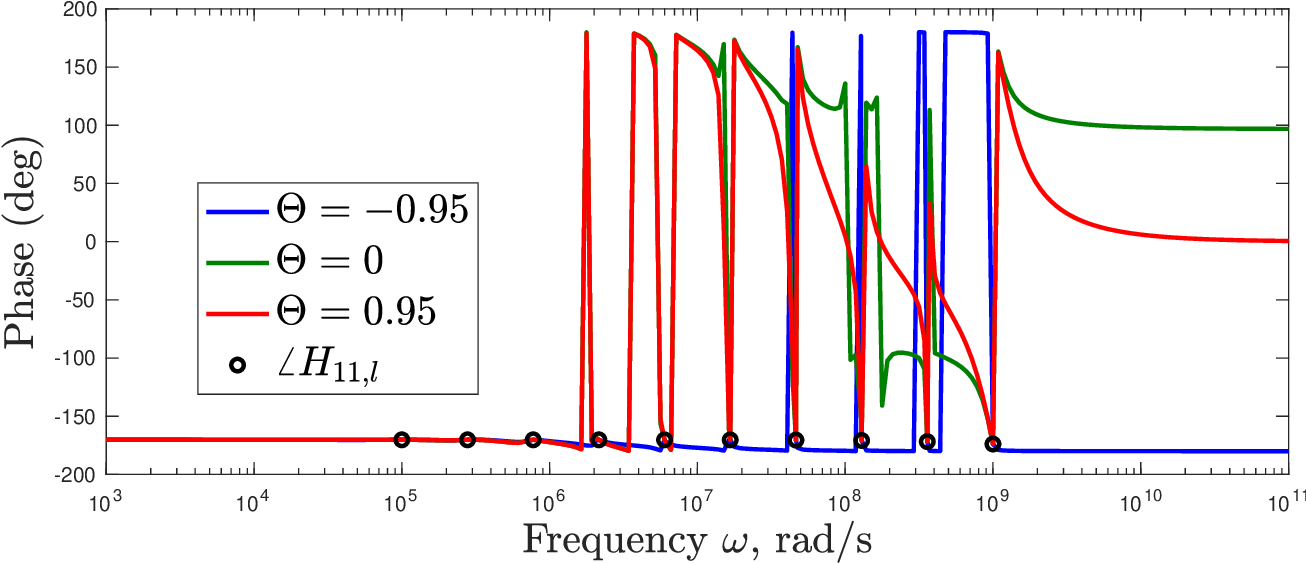}
  \caption{}  \label{fig:ratio}
\end{subfigure}
  \begin{subfigure}{\columnwidth}
  \includegraphics[width=0.9\columnwidth]{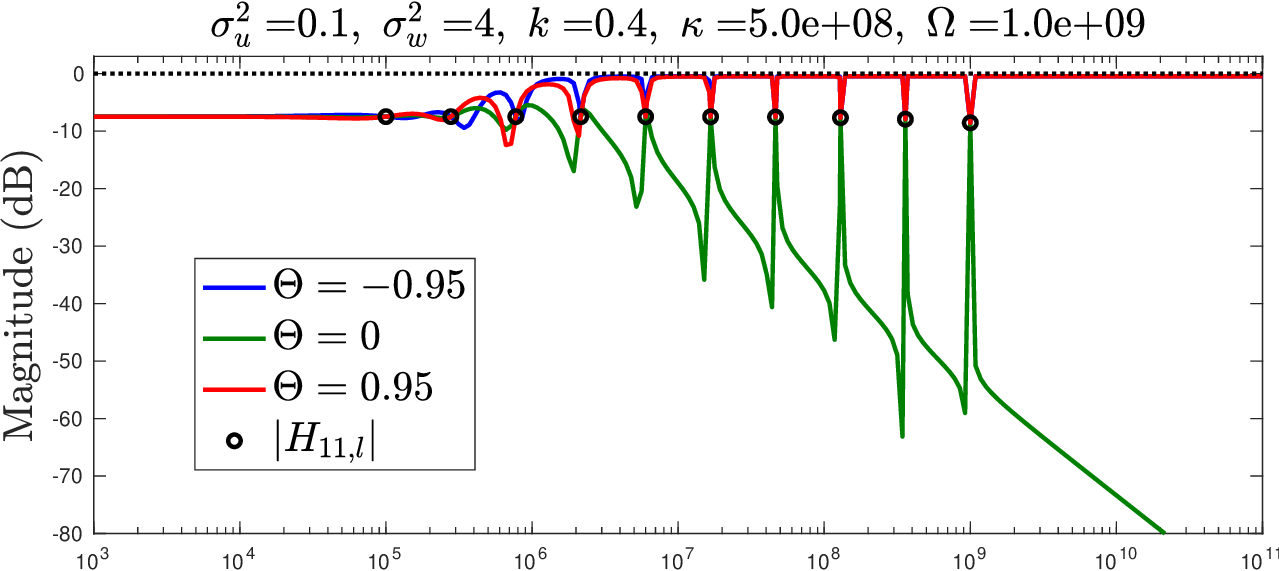}
  \includegraphics[width=0.9\columnwidth]{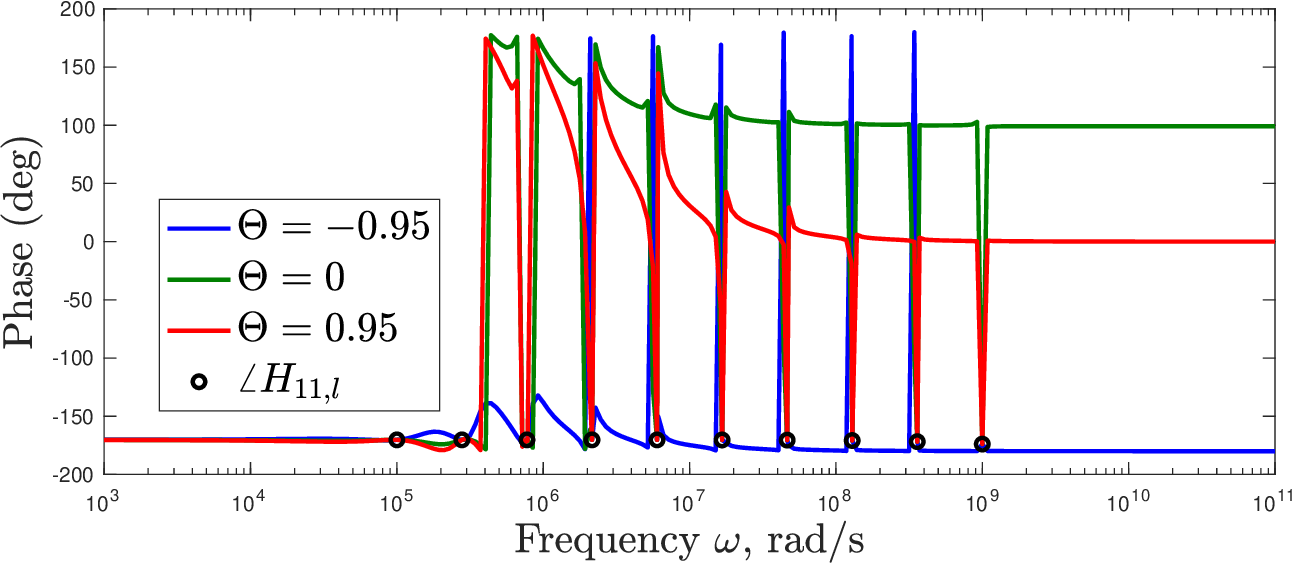}
  \caption{}  \label{fig:ratio.4}
\end{subfigure}
\caption{The Bode plots of $H_{11}(s)$ for $\Theta=-0.95,0$ and $0.95$. The
    circles indicate the magnitude and phase of $H_{11,l}$ obtained from
    the optimization problem~(\ref{eq:71.LMI.l})--(\ref{eq:14.l}): (a)
    $\sigma_w^2=0.2$; (b) $\sigma_w^2=4$.}\label{fig:ratio.all}
\end{figure}

Fig.~\ref{fig:ratio} confirms that all three computed $H_{11}(i\omega)$
agree at the grid frequencies.
The sharp peaks in the graphs occur at the grid point frequencies
  $\omega_l$ which were used in the
LMIs~(\ref{eq:13.l}),~(\ref{eq:14.l}).  These peaks 
occur because the transfer functions $W_{ij}(s)$ which parameterize the
solution have poles at $-\tau+i\omega_l$; see~(\ref{eq:101}). When
$\omega_l\gg \tau$, these poles are quite close to the grid points
$i\omega_l$ on the imaginary axis at which $H_{11}(i\omega)$ was evaluated.

Other than at the control frequencies $\omega_l$, all three graphs of
$P_e(i\omega, 
H_{11})$ deviate from the optimal value $\tilde\gamma^2$ of
  the problem (\ref{eq:71.LMI.l})--(\ref{eq:14.l}). This is expected
since the algorithm optimizes the PSD of the error at selected frequency points
only. It is worth noting that away from
the grid frequencies, $P_e(i\omega, H_{11})$ varies considerably, depending
on $\Theta$. When we let $\Theta=-0.95$ 
in~(\ref{eq:58}), interpolation led to a substantially improved error power
spectrum density, in comparison with the power spectrum
  density $P_{y-u}(i\omega)$ of the difference $y-u$. However, when
$\Theta=0.95$, the error power spectrum density deviated considerably from
the value $\tilde\gamma^2$, and was relatively close to
$P_{y-u}(i\omega)$. Nevertheless, the observed reduction of the
  error PSD using $\Theta=-0.95$ and $\Theta=0$ indicates that there is 
room for further optimization of 
$P_e(i\omega,H_{11})$ over the parameter $\Theta(s)$. Simulations performed
with other values of $\sigma_w^2$ (e.g., see Fig.~\ref{fig:psds.sig2w4})
confirmed this finding. This interesting problem will be 
addressed in future research.  

Another interesting observation is that with the selected
  parameters, the optimization problem~(\ref{eq:71.LMI.l})-(\ref{eq:14.l})
  produced a set of points $H_{11,l}$ that were quite close
  to the boundary of the set $|H_{11}(i\omega)|\le 1$. As a result, while
  all three interpolants satisfied the constraint~(\ref{eq:19}) 
  of Problem~\ref{P1a} required for physical realizability of the filter,
  they do so with a rather small margin; see Fig.~\ref{fig:ratio}. 
  The intuition gained in the previous section suggests that when the noise
  intensity is sufficiently large, the parameter $H_{11}$ of the filter
  should reduce away from the boundary of the constraint set, in an attempt
  to reduce the contribution of the noise field to the output of the
  equalizer. This is confirmed in Fig.~\ref{fig:ratio.4}, which illustrates
  the results of interpolation when $\sigma_w^2=4$. This  time all obtained 
  $H_{11,l}$ have magnitude of order of 0.4. The corresponding value
  $\tilde\gamma^2=1.8122$.

  It is worth noting that for both values of $\sigma_w^2$, letting
  $\Theta=0$ led to $H_{11}(i\omega)$ vanishing at large
    $\omega$. When 
  $\sigma_w^2=0.2$, the filter with $H_{11}(s)$ that vanishes as
  $\omega\to\infty$ is not mean-square optimal. However, when
  $\sigma_w^2=4$, the equalizer with $\Theta=0$ is the best
    out of the three in terms of performance. In this equalizer, the gain  
  $H_{12}(i\omega)$ dominates $H_{11}(i\omega)$ as $\omega\to\infty$. An
  explanation to this is 
  that when the intensity of the channel noise field becomes very large,
  from the view point of optimizing the mean-square error, it becomes
  advantageous to block high frequency components of the channel output
  altogether and transfer the filter environment field as the filter
  output. Such a filtering 
  strategy may not be beneficial when the information accuracy of the
  system is important --- despite the signal-to-noise ratio is low, the
  noisy channel output still carries some information about its input,
  while the filter environment does not carry such information. Therefore,
  an interesting problem for future research is to find a trade-off between
  mean-square accuracy and information accuracy of coherent equalizers,
  similar to the problem considered recently for classical Kalman-Bucy
  filters~\cite{TZUS1}.

\section{Conclusions}\label{Conclusions}
The paper has introduced a quantum counterpart of the classical equalization
problem. The discussion is focused on passive quantum
channels and passive quantum filters, and is motivated by the
utility and the ease of implementation of passive quantum
systems~\cite{Nurdin-2010,NY-2017}. 

Different from the previous work on developing coherent Wiener
  and Kalman filters, we posed the
quantum equalization problem in the same vein as the classical 
$H_\infty$ filtering problem. However, instead of the
  disturbance-to-error transfer function, we considered the PSD of the
  difference between the input field of the quantum communication channel
  and the output field of the equalizer as the measure of the equalizer
  performance. Accordingly, the filter was 
  sought to guarantee that the maximum eigenvalue of the error PSD was
  below a prescribed threshold. The requirement that such
filter must be physically realizable, adds a constraint on the filter. 
 
We have shown that this problem reduces to a constrained
optimization with respect to one of the blocks of equalizer's transfer function
matrix. Using the $J$-spectral factorization technique, we have developed a
convenient parameterization of the class of suboptimal filters similar to
the Youla parameterization of the class of stabilizing controllers.   

Also, this auxiliary problem was cast as a semidefinite program subject to
frequency-dependent   
linear matrix inequality constraints, and a
tractable constraint relaxation was proposed involving constraints
over a discrete set of frequencies. In addition, the Nevanlinna-Pick
interpolation technique was employed to ensure that the
solution to the relaxed problem yields a physically realizable filter. 
A set of all interpolating filters was also obtained. In principle,
coherent filters obtained this way are not guaranteed to yield an improved
mean-square performance over the entire interval of
frequencies. Therefore it is 
interesting to attempt to further improve performance, e.g., by 
minimizing the error power spectrum density over
the set of interpolating filters. 
Another possible direction for future research is to find a
  trade-off between mean-square accuracy and information accuracy of
  coherent equalizers.

The paper gives two examples of equalization of single mode channels. 
One of them comprises a static quantum system as a channel, and another one
includes a quantum optical cavity. These examples demonstrate that coherent
equalizers can be effective in improving the mean-square accuracy of the
channel. We also showed that in the static case, passive
equalizers are especially beneficial when the intensity of the
  thermal noise from the channel environment exceeds certain
  threshold. A linear matrix inequality condition has been
  introduced to predict such a threshold in a general case.

\newcommand{\noopsort}[1]{} \newcommand{\printfirst}[2]{#1}
  \newcommand{\singleletter}[1]{#1} \newcommand{\switchargs}[2]{#2#1}

\end{document}